\documentclass{emulateapj}

\usepackage{epsfig}
\usepackage{amsmath}
\usepackage{rotating}
\usepackage{natbib}
\usepackage{enumerate}
\usepackage{multirow}
\usepackage{array}
\usepackage{appendix}
\usepackage{comment}

\def\plotonesc#1{\centering \leavevmode
\includegraphics[clip=, width=1.70\columnwidth]{#1}}
\def\plotonescTwo#1{\centering \leavevmode
\includegraphics[clip=, width=1.50\columnwidth]{#1}}
\def\plotoneh#1{\centering \leavevmode
\includegraphics[clip=, width=.95\columnwidth]{#1}}

\def\dbar{{\mathchar'26\mkern-12mu d}} 

\newcommand{\cN}[1]{\mathcal{N}}

\def\gsim{\;\rlap{\lower 2.5pt
 \hbox{$\sim$}}\raise 1.5pt\hbox{$>$}\;}
\def\lsim{\;\rlap{\lower 2.5pt
   \hbox{$\sim$}}\raise 1.5pt\hbox{$<$}\;}

\newcommand{\be}{\begin{equation}} 
\newcommand{\ee}{\end{equation}} 

\newcommand{\bea}{\begin{eqnarray}} 
\newcommand{\eea}{\end{eqnarray}} 

\newcommand{\lla}[1]{\label{#1}}

\newcommand{\brp}[1]{\left( #1 \right)}
\newcommand{\brc}[1]{\left\{ #1 \right\}}

\begin{document}


\title{Thermal Processes Governing Hot-Jupiter Radii}

\author{
David S. Spiegel\altaffilmark{1,2},
Adam Burrows\altaffilmark{2}
}

\affil{$^1$Astrophysics Department, Institute for Advanced Study,
  Princeton, NJ 08540}

\affil{$^2$Department of Astrophysical Sciences, Peyton Hall,
  Princeton University, Princeton, NJ 08544}

\vspace{0.5\baselineskip}

\email{
dave@ias.edu,\\
burrows@astro.princeton.edu,
}

\begin{abstract}
There have been many proposed explanations for the
larger-than-expected radii of some transiting hot Jupiters, including
either stellar or orbital energy deposition deep in the atmosphere or
deep in the interior.  In this paper, we explore the important
influences on hot-Jupiter radius evolution of (\textit{i}) additional
heat sources in the high atmosphere, the deep atmosphere, and deep in
the convective interior; (\textit{ii}) consistent cooling of the deep
interior through the planetary dayside, nightside, and poles;
(\textit{iii}) the degree of heat redistribution to the nightside; and
(\textit{iv}) the presence of an upper atmosphere absorber inferred to
produce anomalously hot upper atmospheres and inversions in some
close-in giant planets. In particular, we compare the radius expansion
effects of atmospheric and deep-interior heating at the same power
levels and derive the power required to achieve a given radius
increase when night-side cooling is incorporated. We find that models
that include consistent day/night cooling are more similar to
isotropically irradiated models when there is more heat redistributed
from the dayside to the nightside. In addition, we consider the
efficacy of ohmic heating in the atmosphere and/or convective interior
in inflating hot Jupiters. Among our conclusions are that (\textit{i})
the most highly irradiated planets cannot stably have $uB \gtrsim 10
{\rm~km~s^{-1} \cdot Gauss}$ over a large fraction of their daysides,
where $u$ is the zonal wind speed and $B$ is the dipolar magnetic
field strength in the atmosphere, and (\textit{ii}) that ohmic heating
cannot in and of itself lead to a runaway in planet radius.
\end{abstract}


\section{Introduction}
\label{sec:intro}
The discovery of the ``hot Jupiter'' class of exoplanets
\citep{mayor+queloz1995, marcy+butler1996} ushered in a variety of
mysteries, of which perhaps the first was how they came to exist,
since planet-formation models did not generally predict Jupiter-mass
objects in few-day orbits\footnote{Although, \citet{struve1952} did
  suggest the possibility that objects such as what are now called
  ``hot Jupiters'' might exist.} \citep{boss1995, lissauer1995}.  When
HD~209458b, the first transiting planet discovered
\citep{henry_et_al2000, charbonneau_et_al2000}, was found to have a
radius $\gsim$30\% larger than Jupiter's, a new mystery was born: why
are some hot Jupiters as inflated as they are
\citep{burrows_et_al2000, fortney+hubbard2004, laughlin_et_al2005a,
  burrows_et_al2007}?  More than a decade later, there were at least
50 transiting planets known with masses greater than a third of
Jupiter's and radii greater than 1.3 times Jupiter's.\footnote{See
  http://exoplanet.eu \citep{schneider_et_al2011},
  http://exoplanets.org \citep{wright_et_al2011},\\ or
  http://exoplanet.hanno-rein.de/ \citep{rein2012}.} The radii of
these objects were initially seen as surprising because theoretical
evolutionary models suggested that, at the inferred ages of the
systems (often more than a billion years old), a hydrogen/helium
gas-giant planet ``should'' have a radius significantly smaller, close
to 1.0~$R_J$, where $R_J$ is Jupiter's radius.\footnote{We note that
  some planets are smaller than expected, which can be explained by
  the presence of either a heavy-element core or heavy elements
  throughout the envelope \citep{guillot2006, burrows_et_al2007}.}
The largest of the hot Jupiters have strikingly large radii, including
the 2.04-$R_J$ HAT-P-32b \citep{hartman_et_al2011}, the 1.99-$R_J$
WASP-17b \citep{anderson_et_al2011b}, the 1.83-$R_J$ HAT-P-33b
\citep{hartman_et_al2011}, the 1.74-$R_J$ WASP-12b
\citep{chan_et_al2011}, and the 1.71-$R_J$ TrES-4
\citep{chan_et_al2011}.  At present, although a number of potential
explanations for these planets' large radii have been suggested, it is
unclear whether a single ``inflating'' mechanism predominates.

\begin{figure*}[t]
\plotonesc{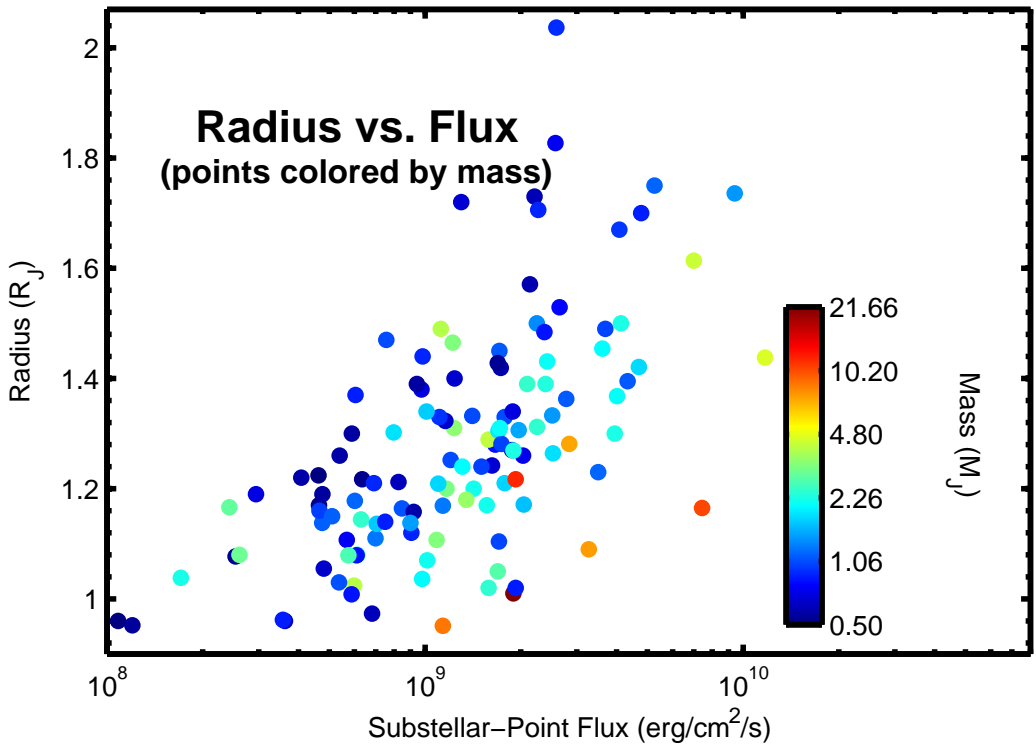}
\caption{Radius vs. incident flux.  For each transiting planet in the
  radius range from 0.9 to 2.07~$R_J$ found on http://exoplanet.eu
  (circa February, 2013), we plot the planet's radius against the
  incident stellar flux at the substellar point; points are colored
  according to planet mass.  The planets with the largest radii tend
  to be highly irradiated and low mass.}
\label{fig:RvF}
\end{figure*}

The processes that have been suggested to explain the surprisingly
large radii include both (\textit{i}) extra power sources in the
interior and (\textit{ii}) enhanced opacity or atmospheric
stratification \citep{burrows_et_al2007, baraffe_et_al2010}.  In
addition, there is a subtle distinction between the radius typically
calculated in evolutionary cooling models (the radius of the
photosphere) and the radius observed in transit; however, this
so-called ``transit radius effect'' \citep{burrows_et_al2003,
  baraffe_et_al2003} increases the apparent radius by no more than
$\sim$5\% relative to the radial photosphere.  Furthermore, the
(``Roche'') shape of an object in a strong tidal field can differ from
spherical, which has a slight effect on the observed transit radius
\citep{Budaj2011}.  Explanations of both category (\textit{i}) and
category (\textit{ii}) tend to rely on the influence of the star and,
therefore, naturally predict that the hottest hot Jupiters (with
incident irradiation more than 10,000 times Jupiter's) should have the
highest probability of being significantly larger than Jupiter.  This
prediction is borne out by observations \citep{laughlin_et_al2011,
  demory+seager2011}, as can be seen in Fig.~\ref{fig:RvF}.

Heating the convective interior of a planet directly increases its
bulk entropy and, hence, its radius \citep{burrows_et_al1994,
  bodenheimer_et_al2001, liu_et_al2008, spiegel+burrows2012,
  marleau+cumming2013}.  Possible sources of heat in the interior
include tidal dissipation of orbital eccentricity
\citep{bodenheimer_et_al2001, Wu_2005_1, Wu_2005_2,
  jackson_et_al2008c, liu_et_al2008, ibgui+burrows2009,
  miller_et_al2009, leconte_et_al2010, ibgui_et_al2010,
  ibgui_et_al2011}, dissipation of thermal tides
\citep{arras+socrates2009a, arras+socrates2009b, arras+socrates2010,
  socrates2013}, dissipation of downwardly propagating gravity waves
\citep{guillot+showman2002, showman+guillot2002}, and ohmic heating
from the dissipation of currents in the partially ionized interior
(\citealt{batygin+stevenson2010, batygin_et_al2011,
  huang+cumming2012}; see also \citealt{perna_et_al2010a}, who
discussed a similar ohmic heating mechanism that might occur in the
radiative portion of the atmosphere).  Since many hot Jupiters have
nearly circular orbits in which there should be little present-day
tidal dissipation, steady-state tidal heating probably cannot
generically explain all inflated radii.  The possibility remains that
a planet might have experienced an episode of tidal circularization in
the recent past, injecting significant energy, increasing its radius,
and effectively ``resetting the clock'' on its evolutionary shrinkage
\citep{ibgui+burrows2009}.  However, since the Kelvin-Helmholtz
timescale of an inflated hot Jupiter planet is on the order of
$\sim$100 million years, far shorter than the ages of the systems,
this sort of explanation requires some fine tuning in the timing and,
therefore, in the initial orbital configuration and efficiency of
tidal dissipation.  While it might explain the radii of some inflated
hot Jupiters, it is not a preferred explanation for the entire
population \citep{leconte_et_al2010, ibgui_et_al2011}.  Indeed,
\citet{ibgui_et_al2011} found via an exhaustive parameter study that
the extremely large radius ($\sim$1.7~$R_J$) of TrES-4, for instance,
cannot be explained simply by a recent episode of tidal dissipation.

Heating the atmosphere of a planet can puff up its outer few scale
heights, but cannot directly explain the radii of many of the larger
hot Jupiters (e.g., those with radii $\gtrsim$1.5~$R_J$ must almost
certainly have inflated interiors).  However, since the atmosphere of
a planet is the conduit through which the convective interior
radiatively loses its heat (entropy), the thermal structure of the
atmosphere critically mediates the evolution of the thermodynamics of
the interior.  The intense stellar irradiation that hot Jupiters
experience can dramatically change the vertical profiles of their
atmospheres, inducing a deep isothermal layer through which
evolutionary cooling is largely effectively stanched
\citep{guillot_et_al1996, burrows_et_al2000, burrows_et_al2003,
  burrows_et_al2006, burrows_et_al2007, chabrier_et_al2004,
  liu_et_al2008, hansen2008, guillot2010}, and significantly modifying
the evolution from that of a less strongly irradiated planet (such as
those of our solar system; \citealt{fortney_et_al2011}) or of widely
separated planets or brown dwarfs \citep{burrows_et_al2001,
  baraffe_et_al2003, marley_et_al2007, fortney_et_al2008b,
  burrows_et_al2011, spiegel+burrows2012, marleau+cumming2013,
  paxton_et_al2013}. The loss of heat from the interior can be further
slowed by increased atmospheric opacity \citep{burrows_et_al2007}, or
by the reduced efficiency of convective heat transport due to
double-diffusive convection \citep{chabrier+baraffe2007,
  leconte+chabrier2012}.  \citet{budaj_et_al2012} examined the
influence on atmosphere structure and on planet cooling of a variety
of physical effects expected in hot Jupiter atmospheres, including
possible extra optical absorbers, the redistribution of day-side heat
to the nightside, and more (although they did not calculate the radius
evolution).  Moreover, extra heating in the atmosphere of a planet,
ultimately driven by the irradiation, can reduce the loss of heat from
the interior.  Sources of extra atmospheric heating include purely
hydrodynamic ones \citep{showman+guillot2002} and magnetohydrodynamic
(MHD) ohmic heating due to atmospheric currents in a partially ionized
atmosphere \citep{perna_et_al2010a, perna_et_al2010b, menou2012,
  rauscher+menou2013}.  The latter process might not only influence
the evolutionary cooling of planets, but also help govern their
present-day weather patterns and wind speeds, and was suggested
simultaneously with, and independently of, the similar mechanism in
the interior introduced by \citeauthor{batygin+stevenson2010}.  Tides
might also provide another source of atmospheric heating, since it is
currently unknown whether the dissipation of tides can deposit a
significant amount of power above the radiative-convective boundary,
which delineates, in our parlance, ``interior'' from ``atmosphere''
\citep{lubow_et_al1997, ogilvie+lin2007, goodman+lackner2009}.

In this paper, we examine the generic character of a variety of
potential mechanisms that have been suggested to explain the inflated
hot Jupiters.  In \S\ref{sec:trends}, we briefly review observed
trends in the dependence of planetary radii and bulk entropy on mass
and incident stellar irradiation.  In \S\ref{sec:basic}, we discuss
how atmospheric heating affects evolutionary cooling and, hence,
radius.  In \S\ref{sec:evolutionary}, we discuss how consistently
coupling the day and night sides of planets in evolutionary cooling
calculations affects the predicted radii given an extra heating
luminosity, or the required extra luminosity to match a given radius
at a given age.  In \S\ref{sec:ohmic}, we explore the effect of
atmospheric ohmic heating, which might be a particularly important
mechanism in planets with large-scale magnetic fields and fast winds.
We present some details of model atmospheres of a variety of hot
Jupiters, including free-electron fraction and conductivity, and
examine how these atmospheric model details might depend on gross
properties of the planets, such as surface gravity and incident
irradiation.  This section concludes by evaluating the stability of
planets against a potential runaway inflation process.  Finally, we
summarize our conclusions in \S\ref{sec:conc}.

\section{Dependence of Radius and Entropy on Mass and Stellar Flux}
\label{sec:trends}

It is instructive to examine the relationship between planetary radius
and various potentially explanatory variables, such as planet mass and
incident stellar flux, among the observed transiting exoplanets.
Here, we simply examine the data for all transiting hot Jupiters,
making no attempt to correct for selection effects.

Figure~\ref{fig:RvF} presents a scatter plot of planet radius versus
incident stellar irradiation.  For each transiting planet found on
http://exoplanet.eu with radius at least 0.9~$R_J$, we color the point
by the planet's mass.  The radii range from our lower cutoff to more
than 2~$R_J$, and the incident fluxes ($F_0$ values) span more than
two decades, from $10^8 {\rm~erg~cm^{-2}~s^{-1}}$ to above $10^{10}
{\rm~erg~cm^{-2}~s^{-1}}$.  The cloud of points exhibit broad scatter,
but several trends are evident.  Perhaps the most obvious of these is
that, even though the color scale is logarithmically spaced in planet
mass, there are still far more blue points than red points, indicating
that lower-mass planets are significantly more numerous in the sample
than higher-mass planets, as has also been seen in some slices of {\it
  Kepler} data \citep{howard_et_al2012}.  Furthermore, the planets
with the largest radii tend to be highly irradiated and relatively low
mass; there are no planets with masses greater than $\sim$5~$M_J$ and
radii greater than 1.29~$R_J$.  The most massive planets do not have
radii much above 1.0~$R_J$.  Finally, there is a general upward slope
associated with the cloud of points, suggesting that $F_0$ could be a
contributing factor to explaining planetary radii.
\citet{demory+seager2011} noted this trend in {\it Kepler} data, as
well, finding that incident stellar flux is positively correlated with
planet radius at irradiation levels above $\sim$$2\times 10^8
{\rm~erg~cm^{-2}~s^{-1}}$.

\begin{figure*}[t]
\plotonescTwo{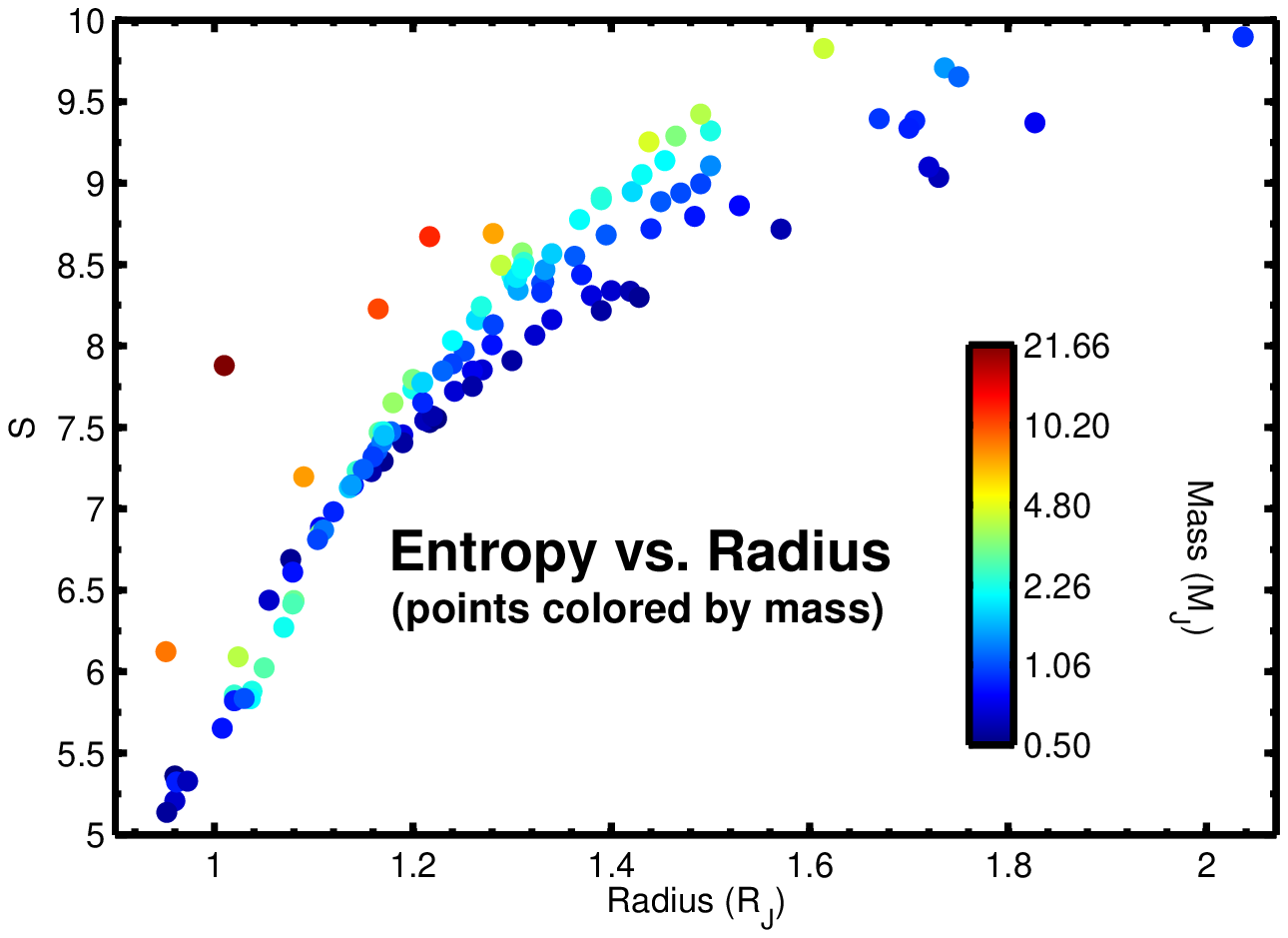}\\
\plotonescTwo{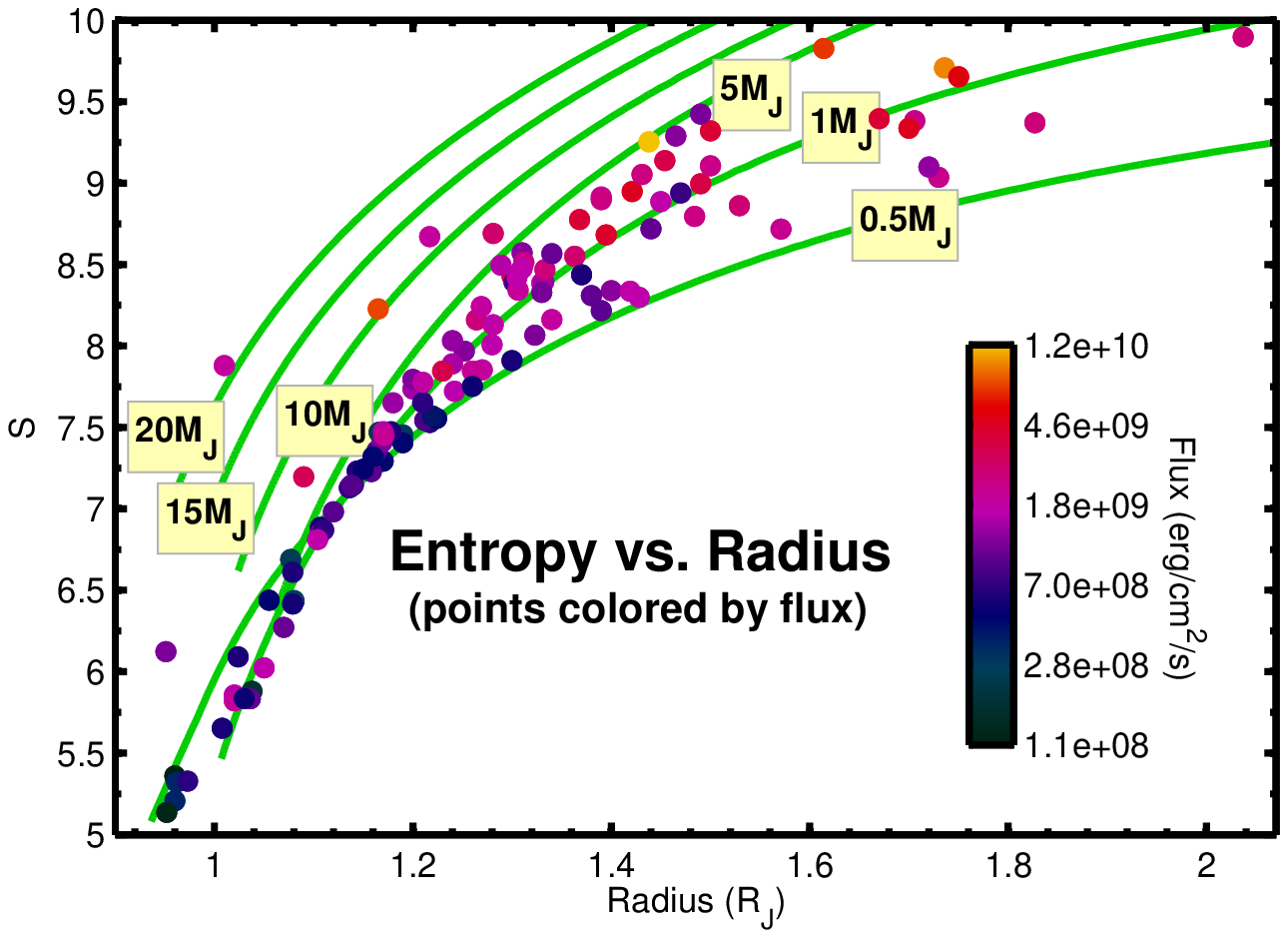}
\caption{Entropy vs. radius, assuming solar composition and no
  heavy-element core.  For each planet in the radius range from 0.9 to
  2.07~$R_J$ on http://exoplanet.eu (in Feburary, 2013), we infer the
  bulk interior specific entropy that corresponds to the listed radius
  and mass.  {\it Top:} The colors of points indicate planet mass.
  The most massive planets do not have radii much above 1.0~$R_J$;
  equivalently, the most inflated planets tend to have lower
  masses. {\it Bottom:} Points are colored according to the incident
  stellar flux.  Contours of constant mass in the radius-specific
  entropy plane are overlaid as green curves.}
\label{fig:SvR}
\end{figure*}

Figure~\ref{fig:SvR} shows inferred planetary specific entropy per
nucleon plotted against planet radius.  For the same sample of planets
shown in Fig.~\ref{fig:RvF}, we compute the bulk interior specific
entropy that corresponds to the listed radius and mass, assuming solar
composition and no heavy-element core.  Both panels show the same
cloud of points; the points are colored in the left panel by mass and
in the right panel by incident stellar flux.  In the right panel,
iso-mass contours are overlaid.  The transiting planets appear to
occupy a fairly narrow swath of the radius-specific entropy plane,
although the true scatter might be slightly different because of
variations in bulk composition and core mass.

\section{The Basic Effect of Atmospheric Heating on Radius}
\label{sec:basic}

\subsection{Modeling Evolutionary Cooling}
\label{ssec:cooling}
At a given mass, the radius of an isolated H/He sphere with a
specified helium fraction depends only on its interior entropy
\citep{zapolsky+salpeter1969}.\footnote{Entropy is a useful diagnostic
  variable because brown dwarfs and giant planets are fully convective
  and, hence, at approximately constant specific entropy throughout
  \citep{burrows+liebert1993}.}  At the same mass and specific
entropy, the radius can be smaller if there is a heavy element core
\citep{guillot2006, burrows_et_al2007} or if either the helium
fraction is larger or there is nonzero metallicity in the bulk
interior \citep{zapolsky+salpeter1969, spiegel_et_al2011a}.  The
thermal state evolves with time as energy flows from the interior to
regions from which the optical depth to infinity is low, at which
point energy eventually radiates to space, leading to a loss of
entropy and a shrinking radius.  This evolution depends on the
atmosphere, since a higher opacity atmosphere will cause a planet to
cool more slowly.

If a planet is not isolated, but rather is irradiated by a nearby
star, energy propagates inward as well as outward.  In such a
situation, the thermal evolution of the interior depends on the {\it
  net} outward radiative flux from the deep interior ($F_{\rm net}$)
through the atmosphere.  The dayside of a highly irradiated planet
tends to have a lower net cooling flux for a given mass and entropy
than an isolated planet has.  The effective temperature ($T_{\rm
  eff}$) of an object is the temperature corresponding to the
bolometric net outward flux ($T_{\rm eff} = \brc{F_{\rm
    net}/\sigma_{\rm SB}}^{1/4}$, where $\sigma_{\rm SB}$ is the
Stefan-Boltzmann constant).  Note that the effective temperature,
which characterizes the difference between the outgoing and incoming
fluxes, should not be confused with the ``equilibrium temperature''
$T_{\rm eq}$, which characterizes just the incoming flux and is
roughly the temperature of the photospheric region of the atmosphere.
In hot Jupiter atmospheres, the approximate flux ratio $(T_{\rm
  eq}/T_{\rm eff})^4$ can be of order $10^4$.  In our evolutionary
calculations, we establish a mapping between surface gravity ($g$),
specific entropy ($S$), and effective temperature by precalculating a
large grid of one-dimensional, non-gray, radiative-convective
atmosphere models with different effective temperatures and surface
gravities and finding the interior adiabat associated with each one,
as described in \citet{burrows_et_al1997}.  One commonly used approach
is to take the downward flux in the atmosphere models to be the
average (over the planet's surface area) of the irradiating flux,
which is $1/4$ of the incident flux at the substellar point (this is
sometimes described as using a ``geometric beaming factor'' $f=1/4$ --
see, e.g., the appendix of \citealt{spiegel+burrows2010}).  We find
the function $s[T_{\rm eff},g]$ and invert it to obtain the cooling
rate $T_{\rm eff}[s,g]$.  This mapping of $(s,g)$ pairs to $T_{\rm
  eff}$ is what we refer to as the ``atmospheric boundary condition.''
(We describe another approach, which couples the cooling through
different portions of the atmosphere --- e.g., the day and night sides
--- in \S\ref{sec:evolutionary}.)

\subsection{The Influence of Atmospheric Heating: Extra Absorber}
\label{ssec:influenceheating_absorber}

The temperature-pressure profile of a planet's atmosphere is
inextricably linked to the planet's rate of cooling.  Radiative
processes that change the vertical thermal structure, therefore,
affect the thermal evolution.

A striking feature that has emerged from studies of the emergent
radiation from hot Jupiters is that many of these objects appear to
have thermal inversions in their upper atmospheres, in which the
temperature increases with height above a relative minimum
\citep{hubeny_et_al2003, burrows_et_al2007c, knutson_et_al2008b,
  fortney_et_al2008, spiegel_et_al2009b, madhusudhan+seager2010b,
  knutson_et_al2010, madhusudhan2012}.  The thermal inversions
presumably result from an enhanced opacity in the short wavelength
part of the spectrum (optical and ultraviolet), above the alkali metal
opacity (mostly sodium and potassium) that some equilibrium chemistry
models have suggested might contribute the bulk of the optical opacity
in hot Jupiter atmospheres \citep{sharp+burrows2007}.  Titanium oxide
(TiO) has been suggested as the source of the extra opacity
\citep{hubeny_et_al2003}, though it is not clear whether hot Jupiter
atmospheres are vigorously enough mixed to bring enough of a heavy,
refractory species to the upper atmosphere where it would need to be
to cause the inversions \citep{spiegel_et_al2009b,
  parmentier_et_al2013}.  Note that ``thermo-resistive'' heating as
suggested by \citet{menou2012b} offers a way to heat the upper
atmosphere without needing to bring a heavy species to great altitude,
although whether it can generically provide sufficient
upper-atmosphere power to explain the inferred upper-atmosphere
heating remains uncertain.

If all else about a model is held constant, an enhanced optical
opacity acts as an anti-greenhouse effect \citep{hubeny_et_al2003},
heating the upper atmosphere and cooling the deeper
atmosphere.\footnote{The greenhouse effect, in contrast, results from
  enhanced infrared opacity and heats the deeper atmosphere while
  cooling the upper atmosphere.}  In particular, at fixed $T_{\rm
  eff}$, increasing the optical opacity decreases the temperature at
depth, in the vicinity of the radiative-convective boundary (see,
e.g., Figs.~1--5 of \citealt{spiegel_et_al2009b}), and, therefore,
decreases the entropy of the convective region.  However, if we
imagine taking a planet that has no thermal inversion and adding an
extra optical absorber so as to create a thermal inversion, the
interior entropy clearly does not suddenly decrease in response to
altered opacity.  Instead, in order to keep the interior entropy
constant, the effective temperature (i.e., the interior cooling rate)
must increase.  One might therefore expect that models with an extra
optical absorber and a thermal inversion would cool faster than those
without these features.  \citet{budaj_et_al2012} show that, at fixed
surface gravity and interior specific entropy, the effective
temperature of a planet increases with the strength of an extra
optical absorber.  Here, we calculate the influence that this process
has on thermal and radius evolution.

\begin{figure*}[t]
\plotonescTwo{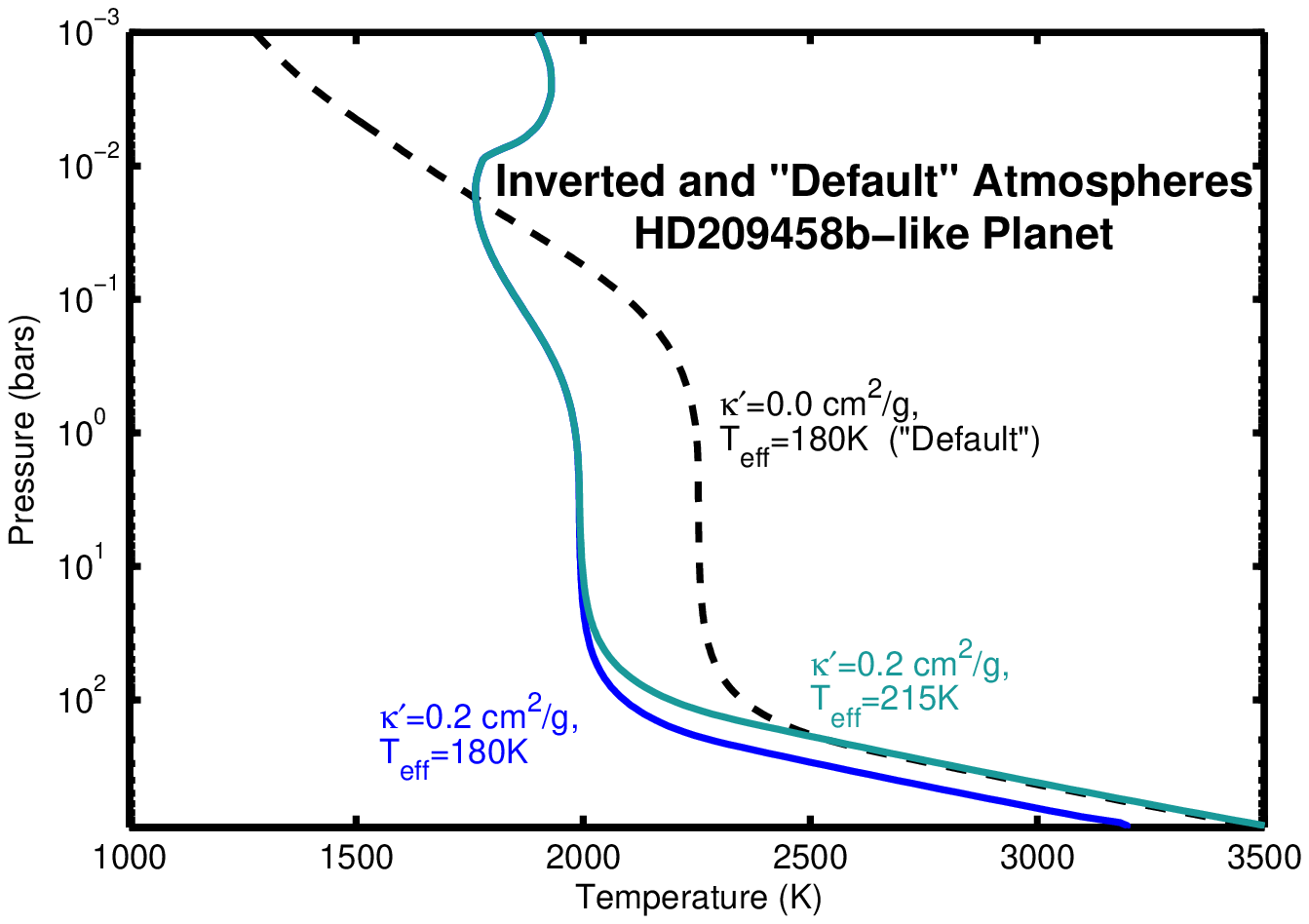}\\
\plotonescTwo{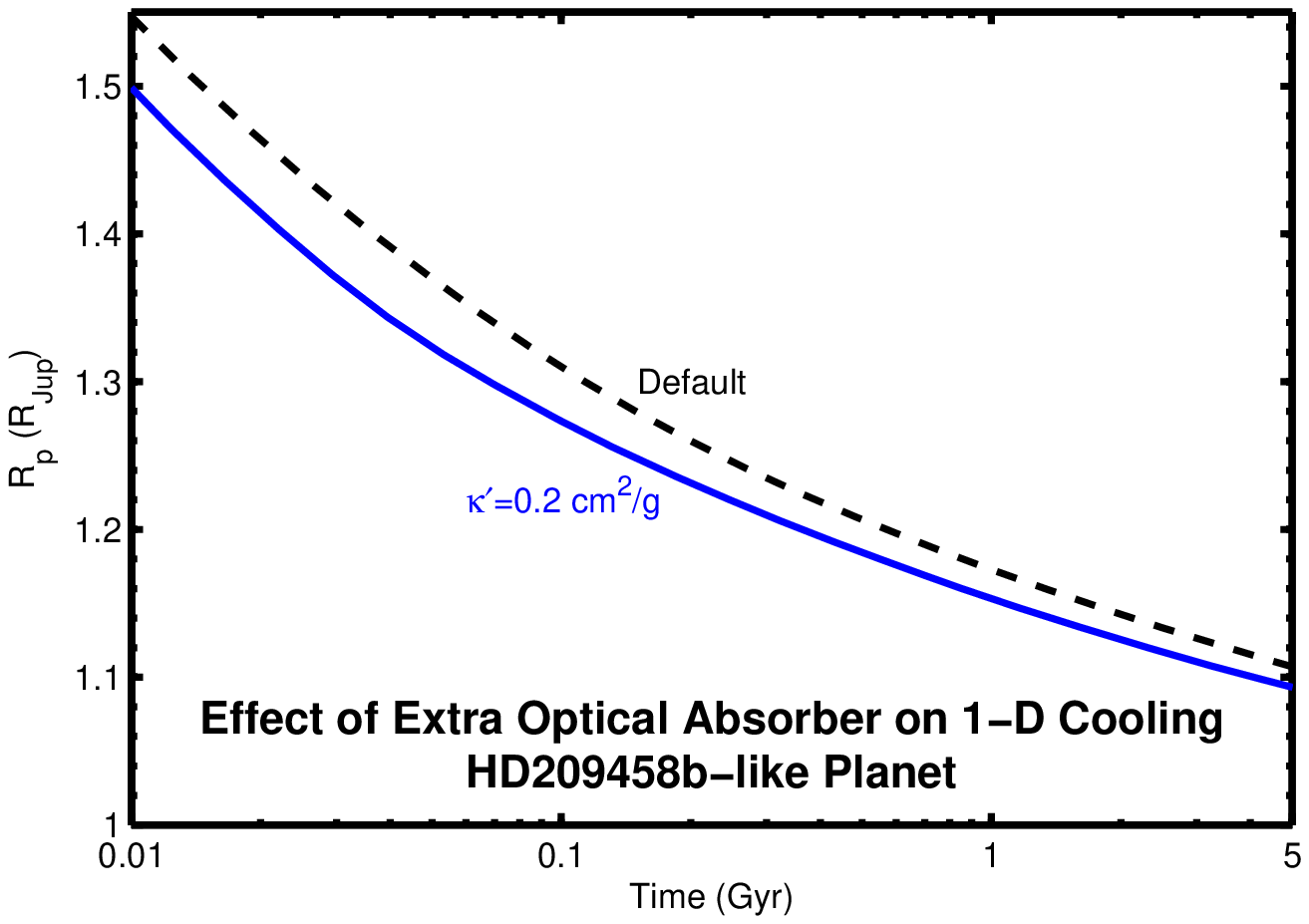}
\caption{Atmosphere and evolution models of HD~209458b-like planet,
  with and without extra optical absorber.  {\it Top:} Comparison of
  atmospheres at constant $g$ and constant $T_{\rm eff}$ (black and
  blue) or constant interior entropy (black and aqua): These
  atmosphere models all have $\log_{10} g = 3.0$ where $g$ is measured
  in cgs units (similar to HD~209458b).  At constant $T_{\rm eff}$
  (dashed black curve and blue curve have $T_{\rm eff} = 180$~K), an
  extra optical absorber ($\kappa'$ --- see, e.g.,
  \citealt{spiegel_et_al2009b}) heats the upper atmosphere and cools
  the lower atmosphere.  At constant interior specific entropy (dashed
  black curve and aqua curve have the same interior adiabat specific
  entropy), the extra optical absorber increases $T_{\rm eff}$ by
  $\sim$20\% and, therefore, doubles the net cooling luminosity. {\it
    Bottom:} Comparison of radius evolution at constant planet mass:
  Two radius evolution trajectories are shown, both of which
  correspond to model planets with a mass of 0.69~$M_J$, where $M_J =
  1.899\times 10^{30} {\rm~g}$ is the mass of Jupiter.  Comparing the
  two radius evolution curves shows that, as expected, in a 1-D
  cooling model, the planet with an extra optical absorber and a
  thermal inversion (blue curve) cools faster and its radius shrinks
  faster than the planet with a ``default'' (noninverted) atmosphere
  (black dashed curve).}
\label{fig:kappa_effect}
\end{figure*}

Figure~\ref{fig:kappa_effect} displays how one-dimensional models
indeed exhibit this behavior.  The top panel of
Fig.~\ref{fig:kappa_effect} shows the influence of adding an extra
optical absorber.  At fixed $T_{\rm eff}$, the deep atmosphere is
cooled relative to an atmosphere without the extra absorber
(represented as $\kappa'$ in the figure).  If the effective
temperature is allowed to vary so as to maintain fixed interior
specific entropy, the model with the extra absorber achieves an
effective temperature nearly 20\% greater (215~K vs. 180~K), implying
a net cooling rate more than twice as great.  This increased cooling
rate is clear in the right panel of Fig.~\ref{fig:kappa_effect}, which
shows the evolution of radius for 1-D model planets with and without
an extra absorber.  The model with the extra absorber shrinks more
rapidly.\footnote{However, the effect of an extra optical absorber
  almost entirely disappears in the context of a ``1+1''-dimensional
  model, as discussed in \S\ref{sec:evolutionary} below.}

\subsection{The Influence of Atmospheric Heating: Below the Photosphere}
\label{ssec:influenceheating_opacity}

Anything that modifies the radiative processes of the atmosphere can
affect evolutionary cooling.  In addition to changes in opacity, as
discussed in \S\ref{ssec:influenceheating_absorber} above, there are
other ways that atmospheres can change the mapping of $S$ and $g$ to
$T_{\rm eff}$.  Adding heat to the atmosphere can affect the interior
entropy and, therefore, the radius.  Consider if some fraction
$\varepsilon$ of the intercepted radiation is converted to another
form of energy (say, mechanical energy) that then dissipates as heat
below the optical photosphere.  At a given mass, effective
temperature, and incident irradiation, a planet where $\varepsilon$ is
greater will have an atmosphere that matches to a higher entropy
adiabat in the convective region, and, therefore, that corresponds to
a higher entropy (and larger radius).

Figure~\ref{fig:heat_constT} illustrates how a power source in the
radiative part of a planet's atmosphere can modify the planet's bulk
entropy (and, hence, the radius) at a given effective temperature.  In
this figure, a range of model planets is displayed, all of which have
the surface gravity and incident stellar irradiation of HD~209458b,
and all have an internal flux temperature of $T_{\rm eff} = 180$~K.
The atmosphere models are calculated with the atmosphere radiative
transfer code {\tt COOLTLUSTY} \citep{hubeny1988, hubeny+lanz1995,
  hubeny_et_al2003, burrows_et_al2006}, and represent solutions to the
radiative transfer equation in which we enforce radiative and chemical
equilibrium.

In the model planets shown in Fig.~\ref{fig:heat_constT}, a small
fraction of the incident irradiation is deposited at various depths in
the atmosphere, with the power spread in a Gaussian distribution over
a region of width 0.5 in log pressure.  The models differ according to
the fraction of incident irradiation that is deposited at depth in the
atmosphere (shown on the abscissa) and according to where in the
atmosphere the power is deposited (shown as different colored curves).
When the power is deposited high in the atmosphere (at pressures below
$\sim$1 bar), it has little influence on the interior adiabat and,
therefore, little influence on the radius.  When power is deposited
deeper, it can have a significant influence on the interior adiabat's
entropy, and a correspondingly large influence on a planet's radius
(at fixed $T_{\rm eff} = 180 {\rm~K}$, with these effects both being
larger when more power is deposited).  The black dashed line shows the
radius ($\approx$1.25~$R_J$) that this model planet has with no extra
atmospheric heating.  Depositing 1\% of the incident power at 10 bars
means that a planet with $T_{\rm eff} = 180$~K has a radius $\sim$5\%
larger (which corresponds to an increment in specific entropy of
$\sim$0.3${~k_B \rm~baryon^{-1}}$, where $k_B$ is Boltzmann's
constant.  If the power is deposited at 30~bars, the planet has a
radius nearly 20\% larger, and if the power is deposited at 100~bars,
the radius is nearly 1.9~$R_J$.

Holding $T_{\rm eff}$ fixed, however, means that the different models
in Fig.~\ref{fig:heat_constT} would correspond to different ages in a
true evolutionary calculation, with the larger radius models
corresponding to younger planets.  This figure demonstrates the
general character of the effect of atmospheric heating on radius, but
a more useful comparison would be to hold constant, not $T_{\rm eff}$,
but rather age.  It turns out that in the context of 1-D models, an
extra power source in the deep atmosphere can result in a dramatically
reduced cooling rate and, hence, significantly inflated radii.  This
effect is muted when night-side cooling is taken into account (see
\S\ref{sec:evolutionary}).

\section{Evolutionary Cooling with Consistent Day/Night Coupling}
\label{sec:evolutionary}

\subsection{The Influence of Non-Isotropic Heating/Cooling}
\label{ssec:noniso}
A gas giant's evolutionary cooling and shrinking are mediated by
details of the three-dimensional atmospheric structure.  Isolated (or
nearly isolated) objects, such as wide-separation planets or brown
dwarfs, may, to a reasonable degree of approximation, be treated as
one dimensional (i.e., as spherically symmetric).  Hot Jupiters, on
the other hand, are expected to be tidally locked to their stars
(although see \citealt{arras+socrates2010}) and, therefore, to
experience a strong, permanent asymmetry in irradiation between the
dayside and the nightside.  As described in \S\ref{ssec:cooling},
treatments in the literature of the thermal evolution of hot Jupiters
have often used one-dimensional models in which the objects are taken
to be irradiated by the global average irradiation, $(1/4)F_0$, where
$F_0$ is the substellar point flux (the ``stellar constant'') for the
planet \citep{guillot_et_al1996, burrows_et_al2000,
  bodenheimer_et_al2001, burrows_et_al2003, baraffe_et_al2003,
  bodenheimer_et_al2003, gu_et_al2003, burrows_et_al2004,
  fortney+hubbard2004, baraffe_et_al2004, chabrier_et_al2004,
  laughlin_et_al2005a, baraffe_et_al2005, baraffe_et_al2006,
  burrows_et_al2007, fortney_et_al2007, marley_et_al2007,
  chabrier+baraffe2007, liu_et_al2008, baraffe_et_al2008,
  ibgui+burrows2009, miller_et_al2009, leconte_et_al2010,
  ibgui_et_al2010, ibgui_et_al2011}.  If horizontal variations in
cooling are large, however, these one-dimensional models might not
accurately capture the character of cooling on a highly-irradiated,
tidally-locked planet \citep{guillot+showman2002, budaj_et_al2012}.

\begin{figure*}[t]
\plotonescTwo{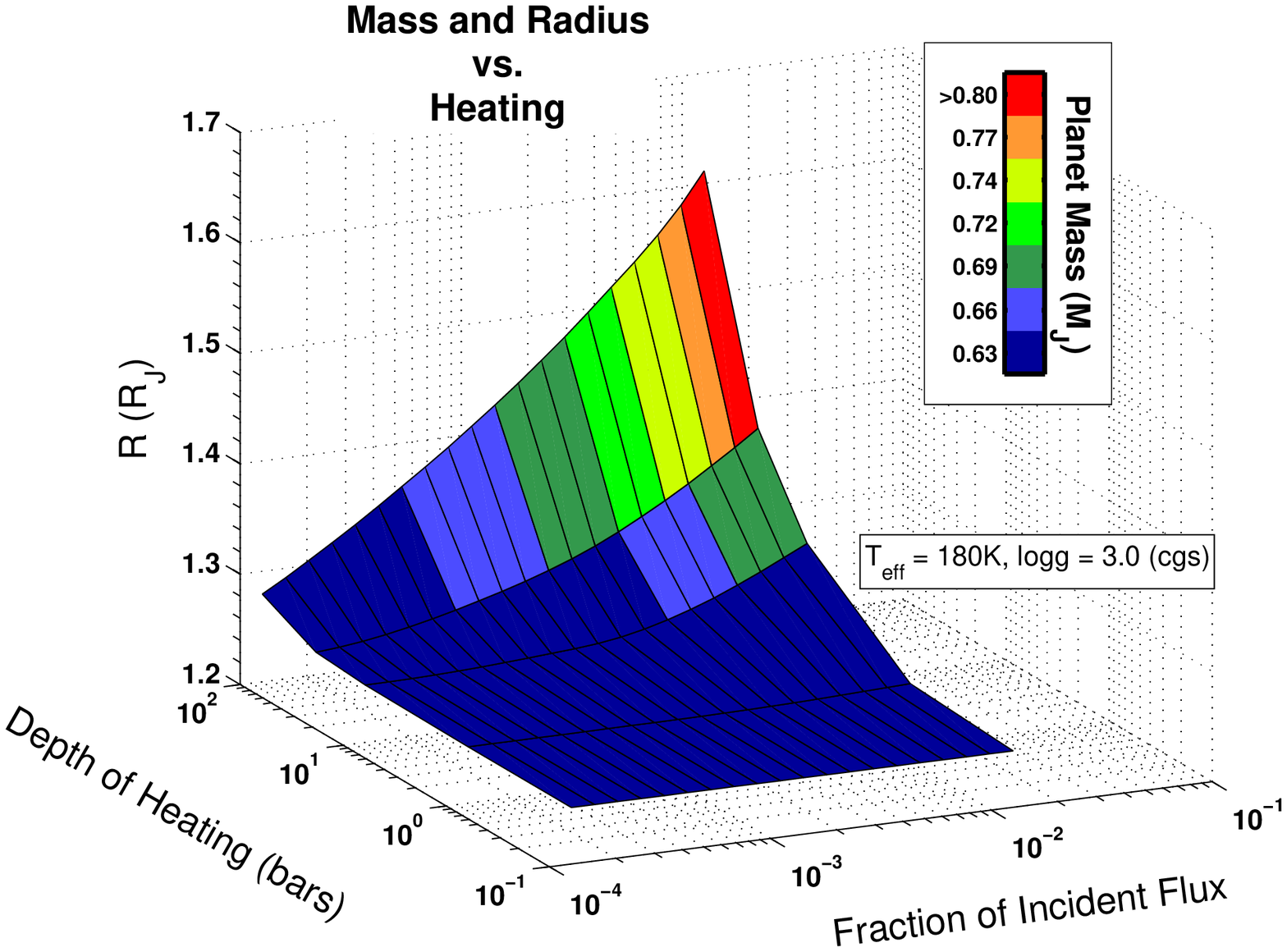}\\
\plotonescTwo{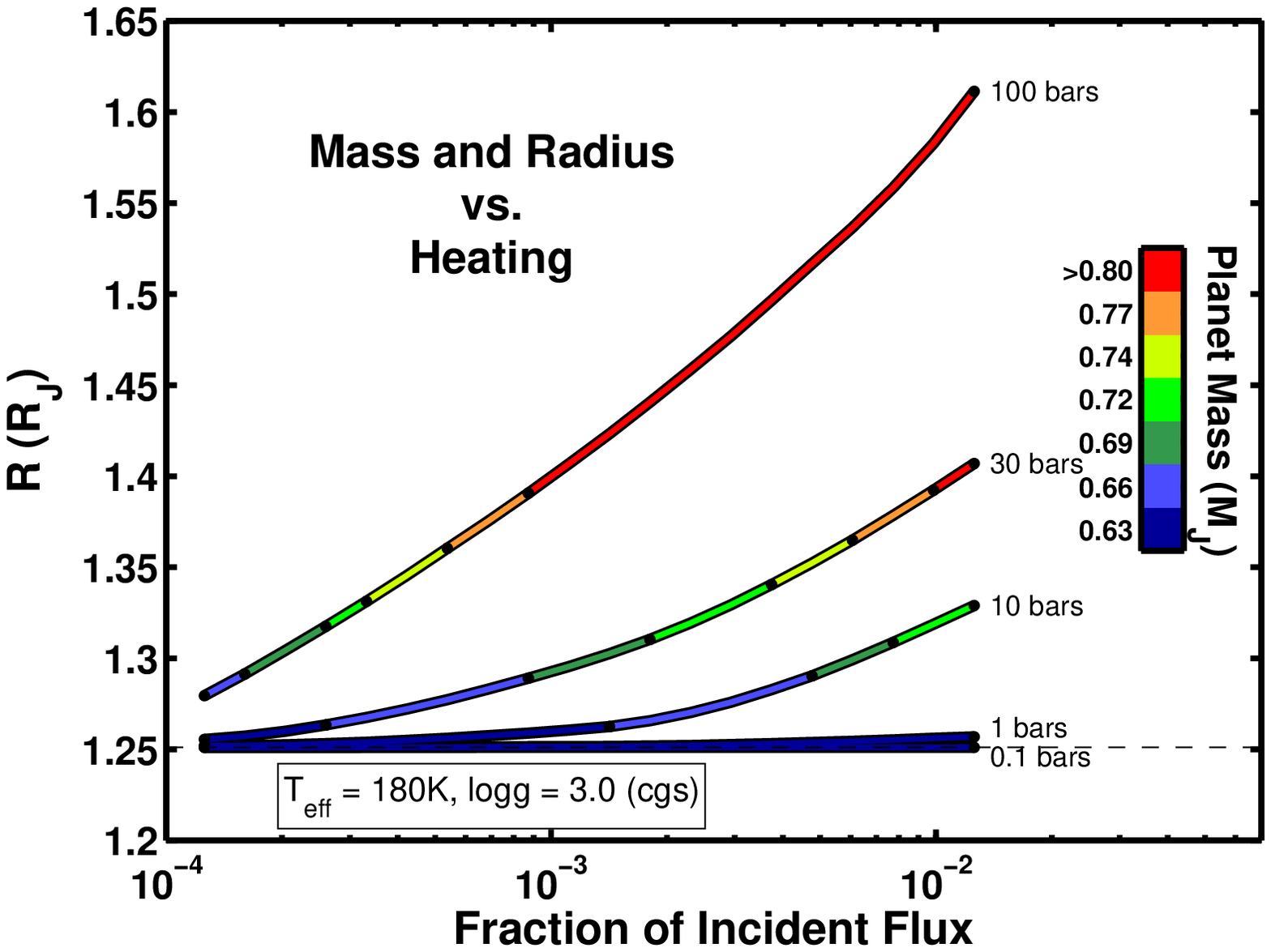}
\caption{Heating at different pressure levels, at constant $T_{\rm
    eff}$.  Each of the model planets shown here has the surface
  gravity ($1.0\times 10^{3} {\rm~dyne~g^{-1}}$) and incident
  irradiation of HD~209458b, and has $T_{\rm eff} = 180$~K.  In each
  model, a small fraction of the incident irradiation (fraction shown
  by the abscissa) is deposited at various depths in the radiative
  portion of the atmosphere, and is spread over a range of pressures
  of log$_{10}$ width 0.5.  {\it Top:} Planet radius (vertical axis)
  and mass (color) are shown as a function of the extra power added to
  the atmosphere and the depth at which it is deposited.  At fixed
  $T_{\rm eff}$ and $g$, greater power and greater depth in the
  atmosphere correspond to larger radius and larger mass. {\it
    Bottom:} The same information as is in the top panel is depicted
  in a different format.  Each curve corresponds to a different depth
  at which power is deposited.  At 100~bars, even a very small
  fraction ($10^{-4}$) of incident power corresponds to an increase of
  a few percent in radius (at constant $T_{rm eff}$ and $g$).  In
  contrast, if the power is deposited at 1~bar, there is hardly any
  influence on radius even if two orders of magnitude more power is
  injected.  The upshot is that greater the atmospheric heating, and
  the greater the depth at which the power is deposited, the higher
  the entropy of the interior adiabat and, therefore, the greater the
  radius (and, at fixed $g$, the higher the planet mass).  The dashed
  black line shows the specific entropy and the radius with no extra
  power.}
\label{fig:heat_constT}
\end{figure*}

Hot Jupiters are presumed to be tidally locked in synchronous
rotation, with one side in permanent day and the other in permanent
night.  The dayside has strong static stability imposed by the steady
injection of energy (and entropy) high in the atmosphere.  The
radiative-convective boundary (RCB) extends deeper into the planet as
the object cools and shrinks, and by an age of $10^9$~yrs or more the
statically stable region on the dayside can extend quite deeply into
the atmosphere, with a RCB at pressures of hundreds of bars to a
kilobar or more \citep{burrows_et_al2003}.  The nightside, with no
irradiation, might have a RCB at a much lower pressure (by several
orders of magnitude), and, therefore, cool more similarly to an
isolated object.  We note that some dynamical atmospheric circulation
models (e.g., \citealt{cooper+showman2005}) find that day-side and
night-side temperatures are fairly similar by a pressure of $\sim$10
bars, although the poles are significantly cooler.  Whether the
relatively cool, less statically stable region is on the nightside or
at the poles is not terribly important; either deviation from an
isotropically cooling planet can cause large changes to the cooling
rate.

One may, therefore, significantly improve upon the models typically
used, by appropriately coupling simple models of the dayside and
nightside together, at consistent interior entropy and surface gravity
\citep{guillot+showman2002, budaj_et_al2012}.  Instead of treating the
entire atmosphere as though it were uniformly irradiated with the
planetary average irradiation, one may produce a ``1+1-dimensional''
model.  First, compute a grid of day-side models, defining a function
$T_{\rm eff}^{\rm day}[s,g]$.  Then, compute a similar grid of
night-side models, defining a function $T_{\rm eff}^{\rm night}[s,g]$.
The total cooling, then, is the sum of the cooling through the dayside
and of that through the nightside:
\be
T_{\rm eff}[s,g]^4 = \frac{T_{\rm eff}^{\rm day}[s,g]^4 + T_{\rm eff}^{\rm night}[s,g]^4}{2} \, ,
\lla{eq:netcool}
\ee
and $T_{\rm eff}[s,g]$ is given by the fourth root of the above
expression.  In equation~(\ref{eq:netcool}), each function must be
evaluated at the same $s$ and $g$: the same $s$ because convection
homogenizes specific entropy, and the same $g$ because the planet is
(very nearly) spherical.  The primary effect of strong irradiation on
the dayside is to create a deep, nearly isothermal layer, through
which the net flux is nearly zero, thereby strongly decreasing
evolutionary cooling (i.e., lowering $T_{\rm eff}^{\rm day}$).  The
nightside and poles lack the dayside's strong irradiation, might
remain fully convective out to much lower pressure levels (bars to
tens of bars), and, thus, might more efficiently transport heat from
the interior to where it may radiate away to space.  As a result, at
equivalent $s$ and $g$, the night-side cooling or polar cooling is
generally much greater than the day-side cooling and, therefore,
$T_{\rm eff}[s,g]^4 \sim (1/2)T_{\rm eff}^{\rm night}[s,g]^4$.  This
can be a large effect that significantly increases the net cooling
relative to what would be found for a hypothetical isotropically
irradiated planet.

The process of horizontal redistribution of day-side atmospheric heat
in hot Jupiter atmospheres has been treated in a number of ways in the
literature (see the Appendix of \citealt{spiegel+burrows2010} for a
review; see also \citealt{cowan+agol2011} and
\citealt{madhusudhan+seager2009}).  The most physically motivated 1-D
parameterization of redistribution is the $P_n$ formalism introduced
by \citet{burrows_et_al2006} and \citet{burrows_et_al2008b}, in which
a fraction $P_n$ (``portion-to-night'') of day-side irradiation is
removed from the day-side atmosphere and deposited on the nightside in
a specified pressure interval (here, $\sim$0.01---0.1~bars, although
the redistribution could in principle happen at other pressure
ranges).  $P_n$ plausibly ranges between 0 (corresponding to no
redistribution) and 0.5 (corresponding to a fully mixed planet where
the nightside receives half the total intercepted power).  In our
models, we use $P_n$ to represent redistribution, removing day-side
flux and adding heat to the nightside.

\begin{figure*}[t]
\plotonesc{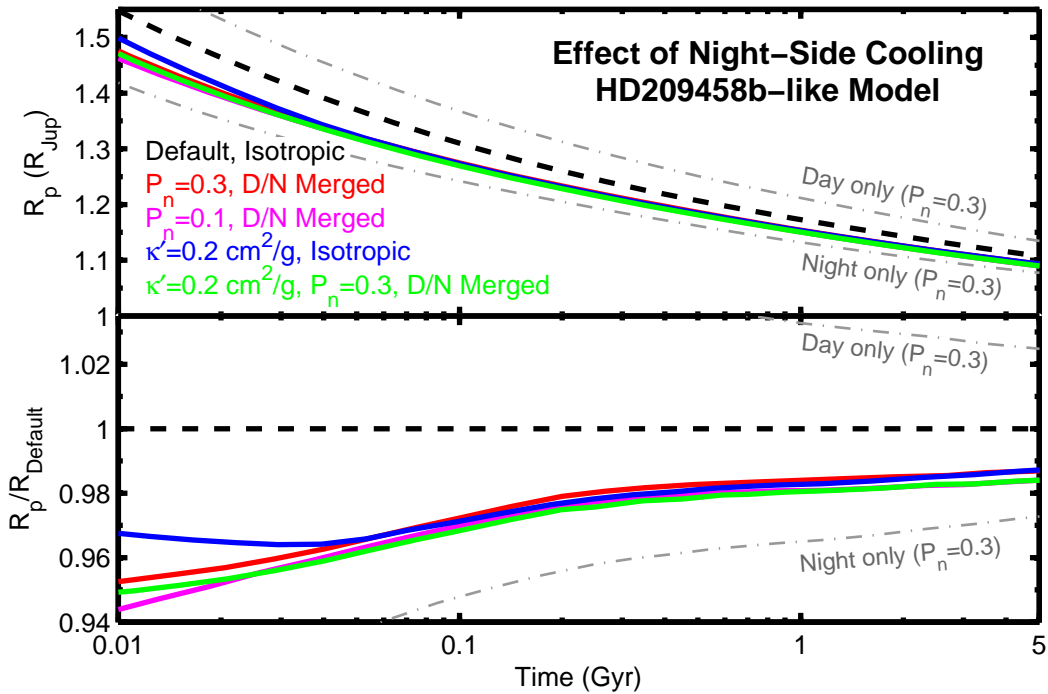}
\caption{Evolution models of an HD~209458b-like planet with and without
  night-side cooling.  Seven radius evolutionary tracks are shown of a
  planet the mass of HD~209458b (0.69~$M_J$), five of which are
  represented with thick lines, corresponding to a 1-D model
  (isotropic irradiation -- black dashed curve), a 1+1-D model with
  $P_n = 0.3$ (red curve), a 1+1-D model with $P_n = 0.1$ (magenta
  curve), a 1-D model with $\kappa'=0.2 {\rm~cm^2~g^{-1}}$ (blue
  curve), and a 1+1-D model with $\kappa'=0.2 {\rm~cm^2~g^{-1}}$ and
  $P_n=0.3$ (green curve).  (The models described as ``D/N Merged''
  are the 1+1-D models with day-side and night-side boundary
  conditions merged consistently.)  For illustrative purposes, we
  include thin gray dashed-dotted lines that indicate evolutionary
  trajectories for hypothetical planets with boundary conditions
  corresponding to day-side only and night-side only boundary
  conditions (i.e., all $4\pi$ steradians are bounded by either the
  dayside or the nightside).  Real planets clearly do not experience
  isotropic day-side or isotropic night-side conditions.  The top and
  bottom curves represent extreme cases that bound the possible range
  of evolutionary tracks for mergers of irradiated and non-irradiated
  boundary conditions (with $P_n=0.3$).  {\it Top:} The evolution of
  the seven models is shown measured in Jupiter radii. {\it Bottom:}
  The evolution of the same models is shown in units of the radius of
  the ``default'' model of the same age.}
\label{fig:night_effect}
\end{figure*}

Figure~\ref{fig:night_effect} shows the evolution of the radius of
HD~209458b-like models with various atmospheric boundary conditions.
Of the seven evolutionary trajectories presented, four represent
isotropic (1-D) models, and three represent day/night-merged (1+1-D)
models.  Two curves are shown that also appear in
Fig.~\ref{fig:kappa_effect}: the ``default'' 1-D model, shown as a
black dashed line; and the $\kappa'=0.2 {\rm~cm^2~g^{-1}}$ model with
a thermally inverted layer created by the extra absorber, shown as a
blue curve.  As previously noted, the effect of the extra absorber is
to increase the cooling rate and reduce the planetary radius at a
given age.

However, when we consider the effect of an extra absorber in a 1+1-D
model with night-side cooling, the influence of nonzero $\kappa'$
nearly disappears.  The day/night-merged models in
Fig.~\ref{fig:night_effect} are bounded by gray dashed-dotted curves
that indicate extreme (and unrealizable) atmosphere conditions.  These
cases consist of isotropic conditions corresponding to day-side only
(upper curve) or night-side only (lower curve) atmospheric conditions.
The night-side only object has zero irradiation and is significantly
heated only by the redistributed heat from the dayside, in this case
corresponding to 30\% of the incident irradiation (the heating from
below corresponding to the net cooling flux of the object pales in
comparison with the redistributed heating from the dayside).  Since
real planets do not experience isotropic day-side or isotropic
night-side conditions, the top and bottom curves simply illustrate
bounding cases for the possible range of evolutionary tracks for
combinations of irradiated and non-irradiated boundary conditions
(with $P_n=0.3$).  The red and magenta curves show 1+1-D evolution
models for $P_n=0.3$ and $P_n=0.1$, respectively.  The green curve
represents a 1+1-D model for $\kappa'=0.2 {\rm~cm^2~g^{-1}}$ and
$P_n=0.3$.  Importantly, the red, magenta, blue, and green curves are
nearly indistinguishable after a few tens of Myr.  In other words,
when night-side cooling is incorporated, all 1+1-D models in
Fig.~\ref{fig:night_effect} cool faster than the default model, and
subtleties such as, in this context, the presence of an optical extra
absorber become insignificant.

\subsection{The Influence of Atmospheric vs. Deep Interior Heating}
\label{ssec:atm_vs_int}

Various mechanisms might deposit at depth an amount of power that is
small in comparison with the irradiating power.  This extra heating
might occur deep in the atmosphere, at pressures of bars to hundreds
of bars, or it might occur in the deep convective interior.

\begin{figure*}[t]
\plotonesc{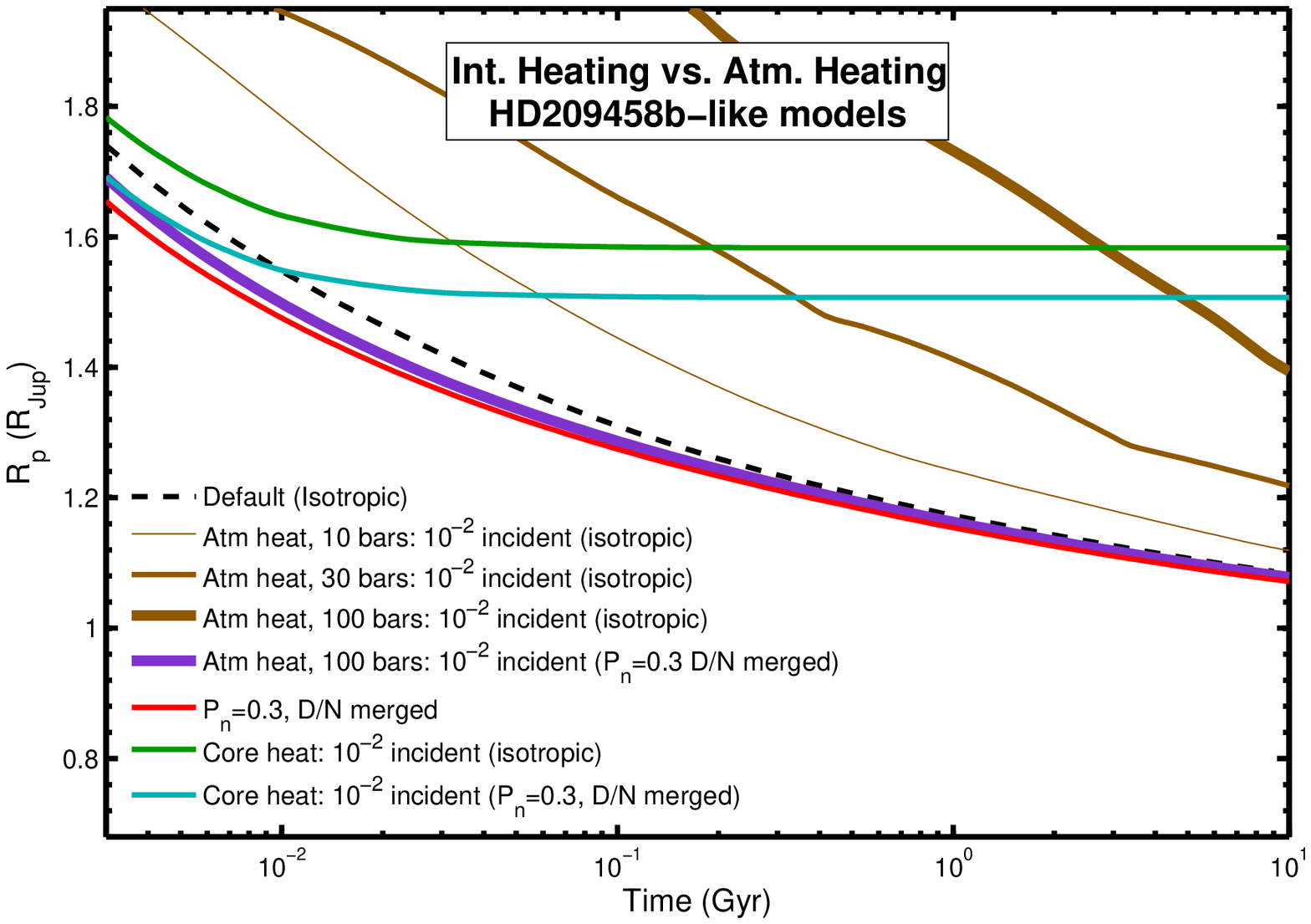}
\caption{Evolution models of an HD~209458b-like planet, with and
  without deep-interior or atmospheric heating.  Eight models are
  presented: the default (isotropic heating) model (black dashed);
  three 1-D models with extra atmospheric heating set to 1\% of the
  incident irradiation, deposited at 10 bars (thin brown), 30 bars
  (medium brown), and 100 bars (thick brown); a 1+1-D model with
  $P_n=0.3$ and with extra heating deposited at 100 bars and an
  unheated nightside (thick purple); a 1+1-D model with no extra
  heating and $P_n=0.3$ (red); a 1-D model with deep-interior heating
  that is 1\% of incident (dark green); and a 1+1-D model with central
  heating that is 1\% of incident and with $P_n=0.3$ (aqua).  In order
  to limit the number of curves for ease of viewing, we present only a
  single purple curve (i.e., we do not display the evolution for when
  the heating is deposited at different pressure levels).  Depositing
  the heat at 10 or 30 bars makes no discernible change to the purple
  curve.  With 1-D models, if extra heating is deposited deep (e.g.,
  100~bars), the radius can be inflated more by heating in the
  atmosphere than by heating in the deep interior until late times
  (several Gyr); however, when night-side cooling is incorporated,
  extra heating in the deep interior inflates the radius more at all
  times later than a few Myr.}
\label{fig:heat_int_atm_effect}
\end{figure*}

Figure~\ref{fig:heat_int_atm_effect} shows radius evolution models of
an HD~209458b-like planet with and without either deep-interior or
atmospheric heating.  Three models demonstrate the influence of extra
isotropic atmospheric heating set to 1\% of the incident irradiation
(brown curves), in which the extra atmospheric power is deposited in a
Gaussian distribution of width 0.5 in log pressure, centered on 10
bars (thin curve), 30 bars (medium curve), and 100 bars (thick curve).
Note that when extra heating is at a depth of tens of bars or more, it
acts to retard evolutionary cooling, in contrast to the effect of
$\kappa'$ --- which is essentially extra heating at millibars or
higher --- which acts to accelerate cooling and shrinking.
Differences in radius between different models generally (but not
always) shrink with time, but after 1~Gyr, the model in which 1\% of
incident irradiation is deposited at 10 bars is $\sim$0.1~$R_J$ larger
than the default model; the model with extra power at 30 bars is more
than 0.2~$R_J$ larger than the default model after 1~Gyr; and the
model with extra power at 100 bars is more than 0.5~$R_J$ larger than
the default model after 1~Gyr, and has a radius larger than 1.7~$R_J$
at this age.  This is consistent with the findings of
\citet{guillot+showman2002}, who argued that depositing 1\% of the
incident irradiation deep in the atmosphere could have an important
radius-inflating effect.

Models in which the dayside's extra atmospheric heating is merged with
night-side cooling exhibit a significantly reduced influence of extra
day-side heating.  This coupled day/night cooling scenario with
day-side heating is represented (for the case of heating centered at
100 bars) by the purple curve.  If the nightside is unheated, the
extra heating on the dayside, which results in such dramatically
increased radii in the 1-D cases, has essentially no influence on the
radius in the 1+1-D case (note the very small difference between the
1+1-D models with no day-side heating and with day-side heating, but
no night-side heating --- the red and purple curves, respectively).

Figure~\ref{fig:heat_int_atm_effect} also presents two examples of
radius evolution with deep-interior heating --- one for a 1-D model
(green) and one for a 1+1-D model (aqua).  The interior heating in
both models is set to 1\% of the incident irradiating power and is
assumed to occur in the center of the planet.  Note that an extra
power source that is deep in the envelope, but not in the center,
might have a complicated influence on the thermal structure of the
object.  Instead of resulting in uniformly higher specific entropy in
the deep interior, such an extra power source might inhibit convection
\citep{wu+lithwick2013}.  We avoid this complexity by assuming that
the extra heat is deposited in the center, but we note that fully
exploring the structural influence of deep envelope heating is a
subject ripe for future work.  At early times, the brown curves (1-D
models with atmospheric heating) have larger radii than the models
with an equivalent amount of central heating.  However, the models
with deep-interior heating reach an inflated asymptotic radius because
in such a model the effective temperatures has a strict
(positive-definite) floor corresponding to the flux implied by the
power source.  In contrast, the asymptotic radius of a model with
extra atmospheric heating is the zero-temperature, fully degenerate
radius --- approximately 1~$R_J$ for objects of roughly Jupiter's mass
--- because the extra heating simply reduces the net cooling rate to a
lower, but still nonzero, value.  The asymptotic radius of the 1-D
model with central heating is roughly $\sim$0.08~$R_J$ larger than
that of the 1+1-D model with central heating and $P_n=0.3$.

\begin{figure*}[t]
\plotonesc{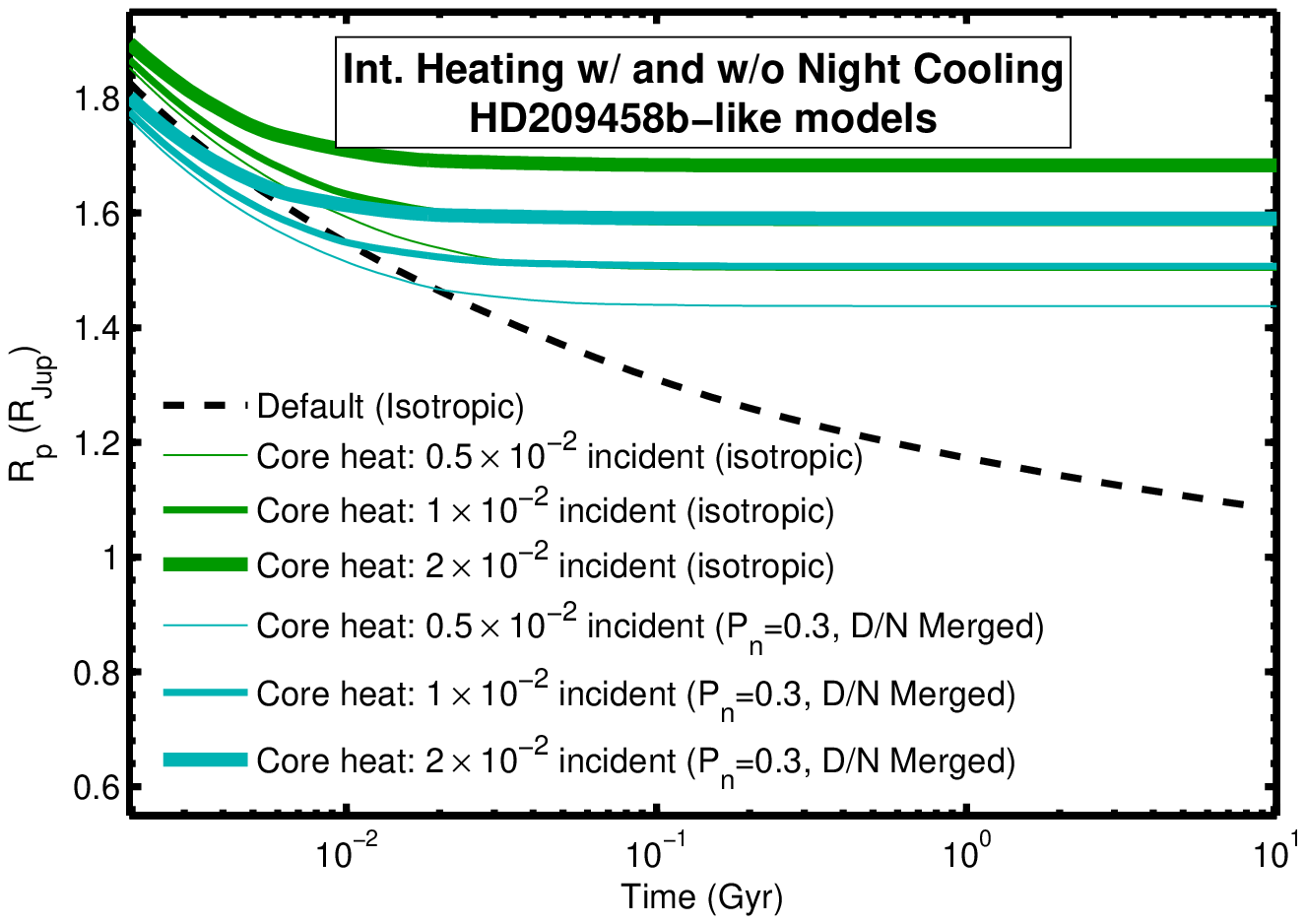}
\caption{Radius evolution models of an HD~209458b-like planet with
  deep-interior heating, with and without night-side cooling.  Seven
  models are presented: the default model, and six models with central
  heating --- 0.5\%, 1\%, and 2\% of incident (thin, medium, and thick
  curves, respectively).  Three of the models with central heating
  have isotropic boundary conditions (dark green), and three have
  1+1-D boundary conditions with $P_n=0.3$.  In this set of models,
  including night-side cooling with $P_n=0.3$ means that the model
  requires a factor of $\sim$2 more interior heating to maintain the
  same radius.}
\label{fig:heat_int_DN}
\end{figure*}

Figure~\ref{fig:heat_int_DN} depicts the radius evolution of models
of an HD~209458b-like planet with a range of levels of central heating
power.  This figure particularly highlights the contrast between 1-D
models (green) and 1+1-D models with $P_n=0.3$ (aqua).  For both the
isotropic model and the one with night-side cooling, increasing the
interior heating power (measured in units of the incident power) from
0.5\% (thin curve) to 1\% (medium curve) results in nearly a 0.1~$R_J$
increase in asymptotic radius, and increasing the power from 1\% to
2\% (thick curve) results in another $\sim$0.1~$R_J$ increase.  For
this set of models, including night-side cooling with $P_n=0.3$
results in a decrement of $\sim$0.1~$R_J$, such that, in the set of
models displayed in Fig.~\ref{fig:heat_int_DN}, achieving the same
asymptotic radius requires twice as much central heating power in the
context of a 1+1-D model with $P_n=0.3$ as with a 1-D isotropic model.

\subsection{The Influence of Redistributive Winds ($P_n$)}
\label{ssec:Pn}
As described in \S\ref{ssec:noniso}, in the absence of winds that
redistribute energy from the dayside of a planet to its nightside, the
nightside would be much cooler than the dayside.  Day-night
temperature differences, however, lead to pressure gradients that
drive redistributive winds (see, e.g., Fig.~3 of
\citealt{burrows_et_al2010}).

\begin{figure*}[t]
\plotonesc{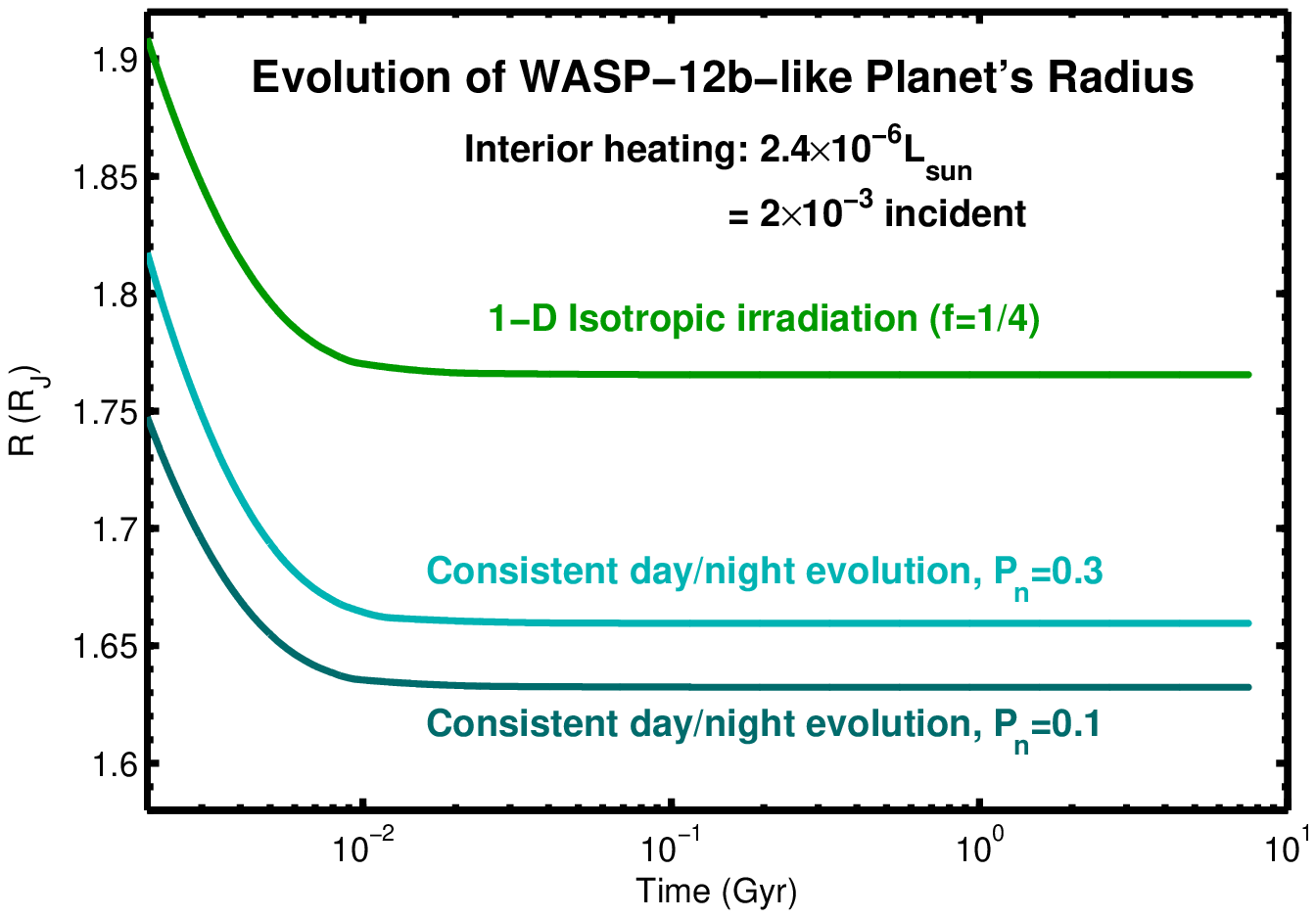}
\caption{Radius of WASP-12b-like planet with interior heating.  The
  evolution of radius is shown for a model planet the mass of WASP-12b
  (1.39~$M_J$) for three different boundary conditions: a default
  ($f=1/4$) model, and two models where the day and night sides are
  consistently coupled, with redistribution parameter $P_n$ set to 0.1
  and to 0.3.  In all three cases, there is an extra interior power
  source equal to $2.4{\times}10^{-6}L_\odot$ (the value needed to
  maintain WASP-12b's radius, according to \citealt{ibgui_et_al2010}).
  The planet cools more through the nightside than through the dayside
  (i.e., the effective temperature of the nightside is greater at
  equivalent entropy and surface gravity), which causes the models
  that incorporate night-side cooling to reach a lower asymptotic
  radius than is reached in the default $f=1/4$ case (which is
  essentially the model of \citealt{ibgui_et_al2010}).  Greater
  redistribution increases the night-side atmospheric temperatures,
  which slows the night-side cooling.  Since the cooling is dominated
  by the night-side cooling, this also reduces the total cooling rate
  and leads to a slightly larger asymptotic radius at equivalent
  central heating power.}
\label{fig:RWASP12b}
\end{figure*}

Figure~\ref{fig:RWASP12b} demonstrates both the effect of
incorporating consistent night-side cooling in an atmosphere model and
the influence of the $P_n$ redistribution parameter.  For variety, in
this figure we show the evolution of a model with the mass and
irradiation of WASP-12b \citep{hebb_et_al2009} instead of HD~209458b.
The model planet is evolved with three different atmospheric boundary
conditions and with the extra interior power that is necessary,
according to \citet{ibgui_et_al2010}, to maintain its current radius
at $\sim$$1\frac{3}{4}$~$R_J$ ($2.4\times 10^{-6}L_\sun$, or 0.2\% of
the incident power).  As above, the source of this extra interior
power is not specified in this model, but it could correspond to tidal
heating or to ohmic heating in the interior.  The three boundary
conditions correspond to a default (isotropic) 1-D model ($f=1/4$),
and two 1+1-D models (with redistribution given by $P_n=0.1$ and
$P_n=0.3$) that couple day-side cooling to night-side cooling in
accordance with Eq.~(\ref{eq:netcool}).  As noted by
\citet{budaj_et_al2012}, redistribution of heat from the day to the
night has the dual effects of cooling the dayside and heating the
nightside.  The former effect slightly increases the dayside's cooling
rate, but the latter effect significantly slows night-side cooling.
The influences of these two effects on radius evolution have opposite
sign, but the (radius-boosting) effect of a warmer nightside is
quantitatively much more significant than the (radius-shrinking)
effect of a cooler dayside.  As a result, the 1+1-D models with
consistent day/night cooling cool more rapidly than does the 1-D
isotropic model, and the model with $P_n=0.3$ asymptotes to a radius
more than 0.1~$R_J$ smaller than the model with isotropic
conditions.\footnote{This same effect was seen previously in
  Figs.~\ref{fig:night_effect}---\ref{fig:heat_int_DN} for
  HD~209458b.}  The isotropic model, unsurprisingly, asymptotes at the
same radius as the analogous model from \citet{ibgui_et_al2010} with
identical interior luminosity.  The model with $P_n=0.3$ has a warmer
nightside than the one with $P_n=0.1$, which leads to a lower net flux
(i.e., a lower effective temperature).  Therefore, since the model
planet's net cooling is dominated by the nightside's effective
temperature, the $P_n=0.3$ model asymptotes at a larger radius (by
0.03~$R_J$) than the $P_n=0.1$ model.

\section{The Effect of Ohmic Heating}
\label{sec:ohmic}

\subsection{Heating in the Atmosphere}
\label{ssec:atmos}
The temperatures and densities that are present in the day-side
atmospheres of highly irradiated planets \citep{fortney_et_al2008,
  spiegel_et_al2009b, madhusudhan+seager2010b, spiegel+burrows2010}
imply a small, but nonnegligible, free electron fraction from the
partial ionization of species with low ionization potential (mostly
atomic potassium and sodium).  During the last decade, a wide variety
of circulation models have predicted that, near the equator and at
depths comparable to the optical and infrared photospheres, hot
Jupiters should have eastward (prograde) wind velocities
($\mathbf{u}$) of $\sim$1$\rm~km~s^{-1}$ or more
\citep{showman+guillot2002, cooper+showman2005, langton+laughlin2008,
  dobbs-dixon+lin2008, menou+rauscher2009, showman_et_al2009,
  heng_et_al2011}.  If planets have large-scale dipolar magnetic
fields ($\mathbf{B}$), zonal winds will carry the free electrons
across field lines, thereby inducing an electric field in response to
the nonzero $\mathbf{u} \times \mathbf{B}$, as pointed out by both
\citet{perna_et_al2010a} and \citet{batygin+stevenson2010}.  The
electric field will drive current loops that could close either in the
atmosphere or in the interior.  Here, we consider the consequences if
currents ohmically dissipate power in the atmosphere.

\begin{figure}[t]
\plotoneh{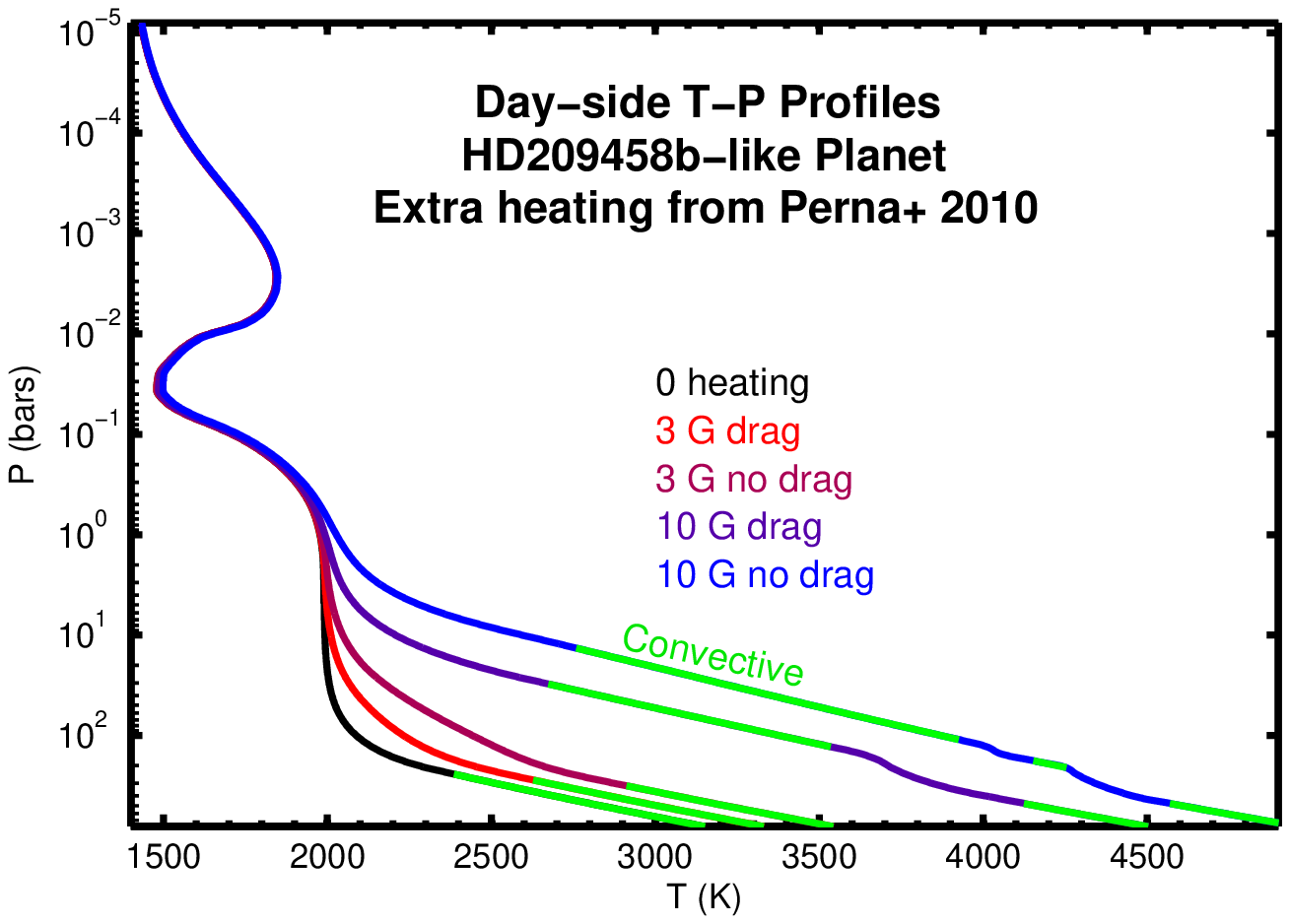}
\caption{Temperature-pressure profiles in the presence of ohmic
  heating.  Five atmospheric profiles are shown (at fixed $T_{\rm
    eff}$) for a planet with the incident irradiation of HD~209458b.
  We adopt the atmospheric heating profiles presented in
  \citet{perna_et_al2010b} for 3-G magnetic fields (with and without
  associated drag on the flow) and for 10-G magnetic fields (with and
  without associated drag).  Stronger magnetic fields lead to greater
  atmospheric heating in the \citet{perna_et_al2010b} models, and
  greater heating causes an atmosphere with a given effective
  temperature to have a base adiabat that is at higher entropy.  Green
  regions are convective (and have uniform entropy) and other regions
  are radiative (and have positive radial entropy gradient).}
\label{fig:thisTP}
\end{figure}

Figure~\ref{fig:thisTP} shows five day-side atmospheric profiles (at
fixed $T_{\rm eff}$) for a planet with the incident stellar
irradiation of HD~209458b.  One profile is for a planet with no extra
heating; the other four have heating profiles taken from
\citet{perna_et_al2010b}, for 3-Gauss (G) magnetic fields (with and
without associated drag on the flow) and for 10-G magnetic fields
(with and without associated drag).  Regions of the atmosphere that
are convective are indicated in green.  In the
\citet{perna_et_al2010b} models, stronger magnetic (and zero drag)
fields lead to greater atmospheric heating.  As seen in
Fig.~\ref{fig:heat_constT}, greater heating (at fixed $T_{\rm eff}$)
leads to greater interior entropy and larger radii.
Figure~\ref{fig:thisTP} illustrates how atmospheric heating at
constant $T_{\rm eff}$ can modify the temperature-pressure profile to
cause the atmosphere to match a higher entropy adiabat --- the same
process that leads to the radius inflation effect of atmospheric
heating seen in Fig.~\ref{fig:heat_int_atm_effect}.

\subsection{Constraints on Ohmic Dissipation}
\label{ssec:ohmic_power}
The ultimate power source of ohmic dissipation is stellar irradiation.
Heating from the star drives winds, which, with a planetary-scale
dipolar magnetic field and with free electrons in the atmosphere, lead
to currents in the atmosphere that ohmically dissipate, dragging the
winds and releasing heat in the atmosphere or in the convective
interior.  The power released by this process is bounded by the power
incident on the planet.

The ohmic heating power per unit volume is $dP_{\rm ohm}/dV =
J^2/\sigma$, where $J$ is the current density and $\sigma$ is the
electrical conductivity.  For a dipolar magnetic field strength $B$,
with zonal wind speed $u$, this can be expressed \citep{menou2012} as
\be
\frac{dP_{\rm ohm}}{dV} = \brp{\frac{u}{c}}^2 B^2 \sigma \, ,
\lla{eq:Pohm}
\ee
where $c$ is the speed of light.  The electrical conductivity can be
written
\be
\sigma = \frac{n_e}{n_n} \frac{e^2/m_e}{\left< Av\right>} \, ,
\lla{eq:conductivity}
\ee
where, $n_e$ and $n_n$ are the number densities of free electrons and
neutrals, respectively, $e$ is the electron charge, $m_e$ is the
electron mass, $A$ is the electron cross section, and, per
\citet{draine_et_al1983}, $\left< Av \right> \sim 10^{-15}(128k_B T /
9 \pi m_e)^{1/2}$.  The ohmic heating density, therefore, is
\bea
\nonumber \frac{dP_{\rm ohm}}{dV} & \sim & 10^3 {\rm~erg~cm^{-3}~s^{-1}} \brp{\frac{u}{1 \rm~km/s}}^{2} \brp{\frac{B}{10 \rm~G}}^2 \\
\lla{eq:PowerDensity} & & \times \brp{\frac{Y_e}{10^{-4}}} \brp{\frac{T}{10^3 \rm~K}}^{-1/2} \, ,
\eea
where the free-electron mixing ratio $Y_e \equiv n_e/n_n$.  The
scaling for $Y_e$ ($10^{-4}$) might seem strangely high, since at
common conditions in hot Jupiter atmospheres most free electrons are
contributed by singly ionized potassium and sodium, whose abundances
(mixing ratios of $\sim$$10^{-7}$ and $\sim$$2\times 10^{-6}$,
respectively, if the composition is solar) are both much lower than
$10^{-4}$.  For the hottest hot Jupiters, however (e.g., WASP-12b and
HAT-P-7b), ionization of hydrogen can contribute up to $\sim$$10^{-4}$
in free-electron mixing ratio (see Fig.~\ref{fig:dayprof}).

If the ohmic heating occurs over a solid angle $\pi$ of the planet's
dayside, and extends vertically over $N$ scale heights $H$, then the
integrated ohmic heating rate is
\bea
\lla{eq:PohmInt} P_{\rm ohm} & \sim & \pi R_p^2 N H \frac{dP_{\rm ohm}}{dV} \\
\nonumber & \sim & \brp{5 \times 10^{30} {\rm \frac{erg}{s}}}  N \brp{\frac{R_p}{R_J}}^2 \brp{\frac{H}{350 \rm~km}} \\
\nonumber &  &  \brp{\frac{u}{1 \rm~km/s}}^{2} \brp{\frac{B}{10 \rm~G}}^2 \brp{\frac{Y_e}{10^{-4}}} \brp{\frac{T}{10^3 \rm~K}}^{-1/2}  .
\eea
The power incident on a planet a distance $a$ from a star of
luminosity $L_*$ is
\be
P_{\rm in} \sim 7 \times 10^{30} {\rm~erg~s^{-1}} \brp{\frac{L_*}{L_\odot}} \brp{\frac{a}{0.01 \rm~AU}}^{-2} \brp{\frac{R_p}{R_J}}^2 .
\lla{eq:Pin}
\ee
Since the ohmic heating clearly cannot exceed the incident stellar
heating, a comparison of equations~(\ref{eq:PohmInt}) and
(\ref{eq:Pin}) shows that there must be feedback between $B$ and $u$,
whereby ohmic dissipation limits wind speeds.  Since the heating power
comes from the dissipation of wind velocities, this point is trivially
true \citep{perna_et_al2010a, perna_et_al2010b, batygin_et_al2011}.
Nevertheless, it is interesting to note that simple energy balance
requires that, if the dipolar planetary magnetic field is of order
10~G, the hottest hot Jupiters {\it cannot} (over a large fraction of
their day-side atmospheres) have zonal winds at the
$\sim$kilometer-per-second level that is typically suggested by
numerical simulations that do not include the magnetic drag effect, a
conclusion that is also implied by the analysis of \citet{menou2012}.

\citet{menou2012} suggests scaling relations for winds in
ohmically-heated hot-Jupiter atmospheres.  Without considering ohmic
dissipation, the horizontal momentum equation leads to an
order-of-magnitude estimate of \citep{showman_et_al2011}
\be
u^2 \sim \frac{k \Delta T_{\rm horiz}}{\mu m_p} \Delta \ln P \, ,
\lla{eq:MomEq}
\ee
where $\Delta T_{\rm horiz}$ is the day-night temperature contrast,
$\mu$ is the mean molecular weight of the atmosphere in units of the
proton mass $m_p$, and $\Delta \ln P$ is the number of scale heights
above a base level of the atmosphere at which the day-side and
night-side atmospheres have the same thermodynamic state.  Note that
equation~(\ref{eq:MomEq}) may be expressed as
\bea
\nonumber u^2 & \sim & \brp{{c_s}_{\rm day}^2 - {c_s}_{\rm night}^2} \Delta \ln P \, ,  \qquad \mbox{or,} \\
 & \sim & \brp{{c_g}_{\rm day}^2[2\pi H] - {c_g}_{\rm night}^2[2\pi H]} \Delta \ln P \, ,
\lla{eq:rewrite}
\eea
where the $c_s$ terms are the sound speeds on the day and night sides,
and the $c_g[2\pi H]$ terms are the gravity wave speeds for wavelength
$2\pi H$ on the day and night sides.  For reasonable values of the
day-night temperature difference, equations~(\ref{eq:MomEq}) and
(\ref{eq:rewrite}) yield zonal wind speeds of order the day-side sound
speed, roughly a few kilometers per second ($\sim$2$\brc{T_{\rm
    day}/1000\rm~K}^{1/2}$~km~s$^{-1}$), which indicates that the
magnetic drag term in equation~(5) of \citet{menou2012} plays an
important role in limiting wind speeds if the magnetic field is of
order several Gauss or more \citep{perna_et_al2010a,
  perna_et_al2010b}.  Note that most red-giant hot Jupiters
\citep{spiegel+madhusudhan2012} probably do not receive enough
irradiation to have a high enough free-electron fraction to experience
non-negligible ohmic heating.

\subsection{Ionization and Heating in the Population of Known Hot Jupiters}
\label{ssec:details}

It is now possible to construct radiative-convective atmosphere
models, guided by observations, of more than 20 transiting exoplanets.
We have published models of over a dozen, including, in order of
increasing incident stellar flux, HD~189733b
\citep{burrows_et_al2008b}, HD~209458b \citep{knutson_et_al2008b,
  burrows_et_al2008b}, TrES-3 \citep{fressin_et_al2010}, HAT-P-7b
\citep{spiegel+burrows2010}, WASP-18b \citep{machalek_et_al2010}, and
WASP-12b \citep{cowan_et_al2012}.  The radii of the members of this
sextet, respectively, are 1.14~$R_J$, 1.35~$R_J$, 1.34~$R_J$,
1.36~$R_J$, 1.11~$R_J$, and 1.83~$R_J$.

\begin{figure*}[t!]
\plotoneh{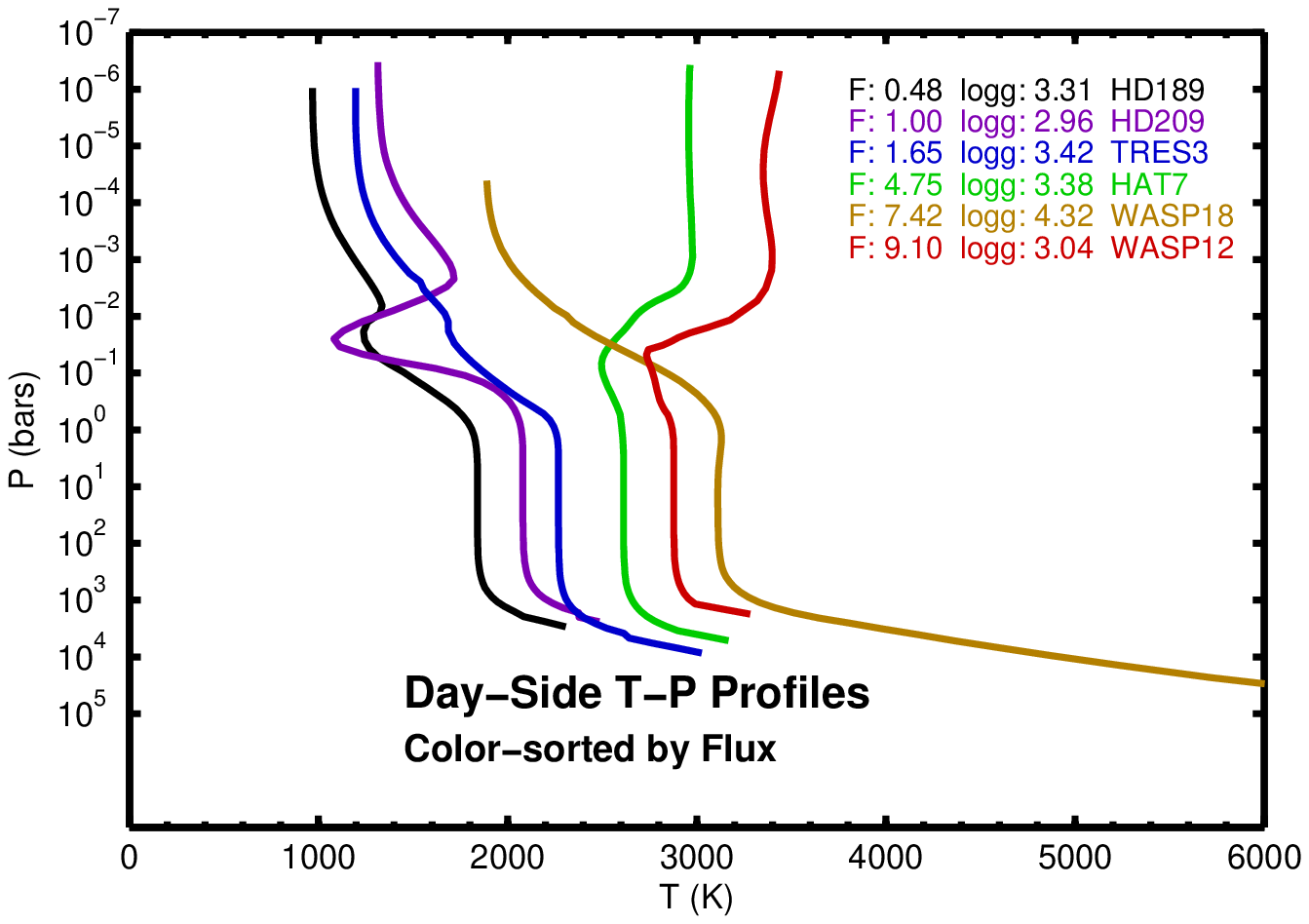}
\plotoneh{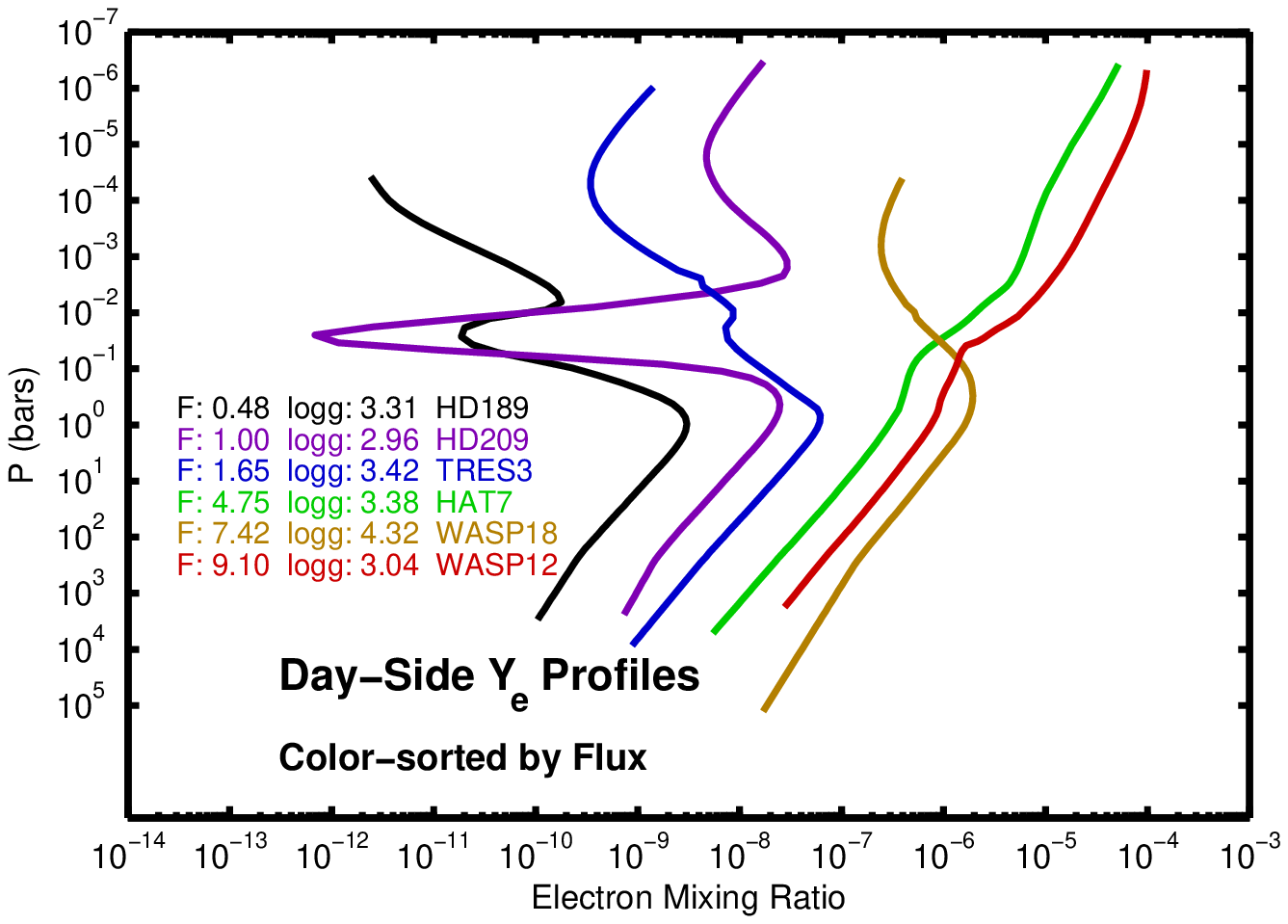}\\
\plotoneh{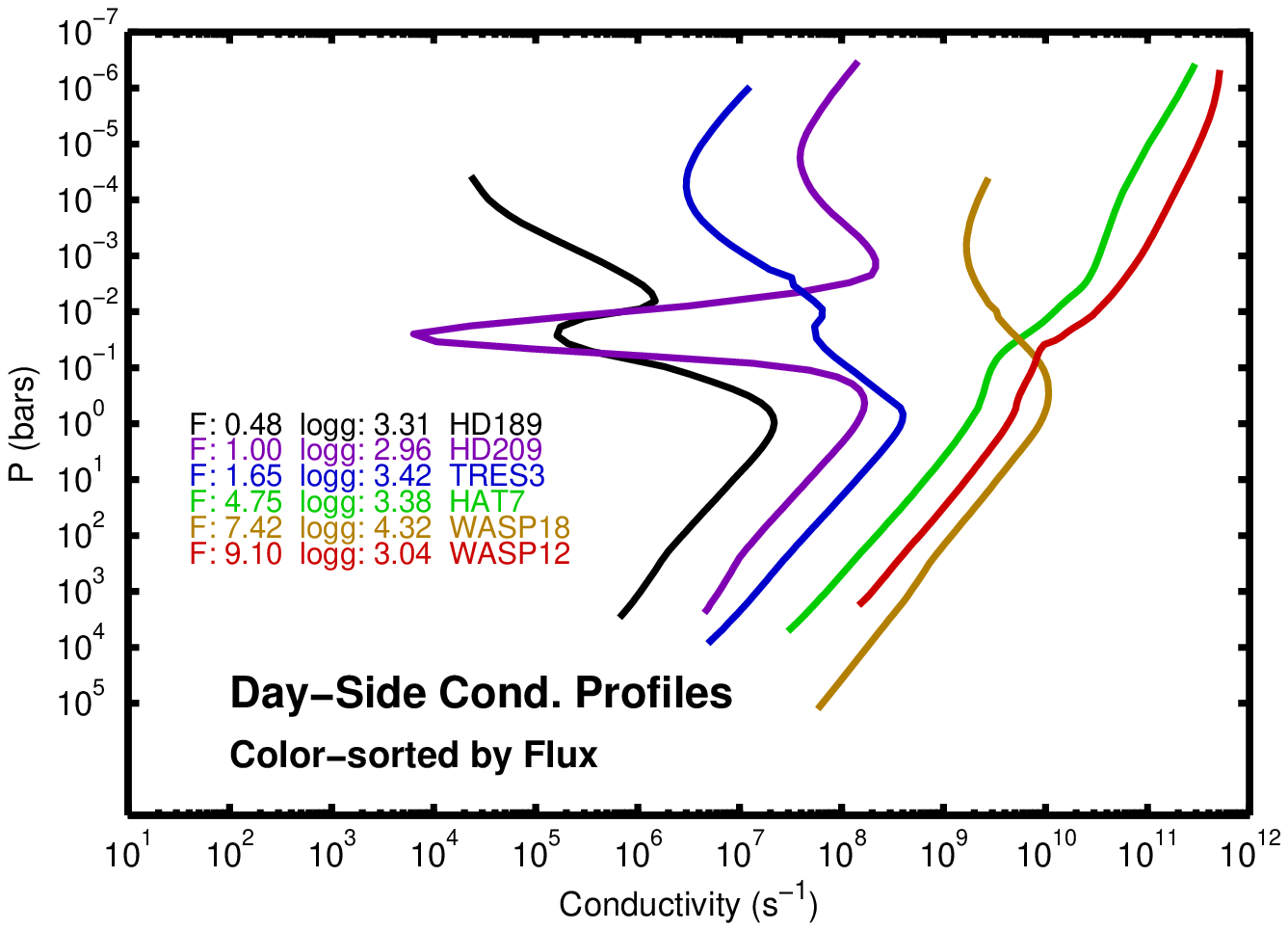}
\plotoneh{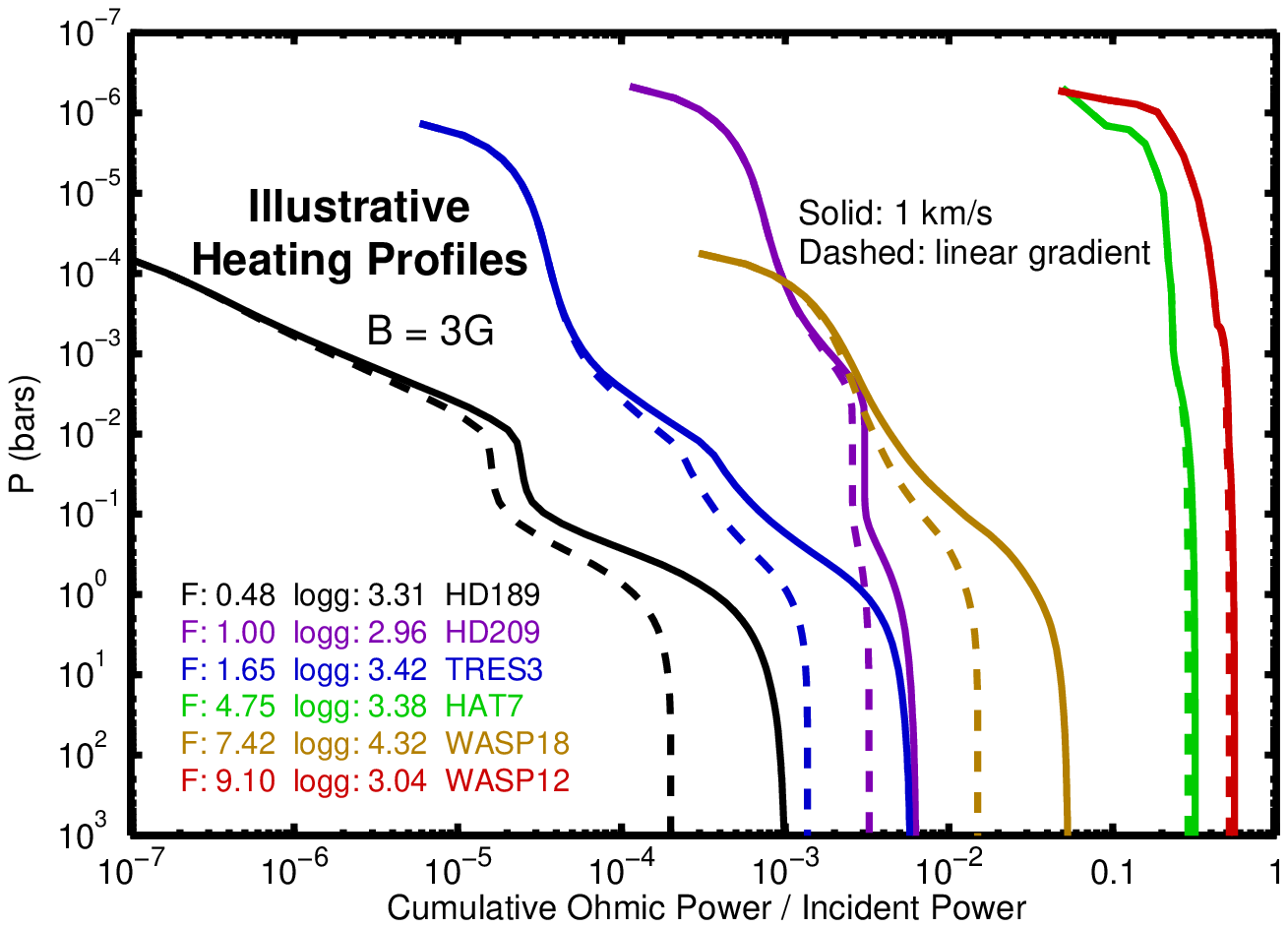}
\caption{Profiles of properties of day-side atmospheres of transiting
  exoplanets.  Each panel shows pressure as the ordinate (decreasing
  upward on a logarithmic scale; i.e., altitude increases upward
  approximately linearly) and some other atmospheric property as the
  abscissa.  The profiles are color-sorted by incident flux, and the
  legend in each panel indicates incident flux (``F:'') in units of
  $10^9 {\rm~erg~cm^{-2}~s^{-1}}$ and the base-10 logarithm of surface
  gravity (``logg:'') in dynes per gram.  {\it Top left:} Temperature
  profiles are shown, including some corresponding to
  thermally-inverted atmospheres (e.g., HD~209458b, WASP-18b, and
  WASP-12b -- although see \citealt{madhusudhan_et_al2011b}).  More
  highly irradiated planets tend to have hotter isothermal layers
  between $\sim$1 and $\sim$300 bars. {\it Top right:} Electron mixing
  ratios are derived from pressure and temperature according to Saha
  equations.  Most of the free electrons in the upper atmosphere come
  from singly ionized sodium and potassium.  All atmospheres shown are
  at most weakly ionized (the maximum ionization shown is
  $\sim$10$^{-4}$, for a model of WASP-12b), but, importantly, the
  ionization is nonzero.  More highly irradiated planets have greater
  partial ionization.  (High in the atmosphere --- at $\sim$microbar
  pressures --- there can be a nonnegligible photoionization component
  to the free electron fraction.)  {\it Bottom left:} Conductivity
  profiles are derived from the free electron number density.  {\it
    Bottom right:} Heating profiles are derived from the conductivity,
  with assumptions about the magnetic field strength (3~Gauss) and
  wind speeds (solid: 1.0~km/s; dashed: 1.0~km/s at pressures lower
  than $10^{-4}$~bars, 0.0~km/s at pressures greater than 300~bars,
  and varying linearly with altitude in between).}
\label{fig:dayprof}
\end{figure*}

Figure~\ref{fig:dayprof} provides illustrative examples of the types
of atmospheric structures that may be found across a range of hot
Jupiter conditions.  We present vertical profiles of (clockwise from
upper left) atmospheric temperature, electron mixing ratio,
conductivity, and ohmic heating for the six planets mentioned above,
for which the incident flux varies from 0.48 to
9.10$\times$10$^9$~erg~cm$^{-2}$~s$^{-1}$, and log$_{10}$[$g$] (in
cgs) ranges from 2.96 to 4.42.  The profiles assume day-side average
conditions.

The top-left panel of Fig.~\ref{fig:dayprof} shows that three of these
models -- HD~209458b, HAT-P-7b, and WASP-12b -- have thermal
inversions in their upper atmospheres, although that of WASP-12b is
somewhat disputed (see \citealt{madhusudhan_et_al2011b},
\citealt{cowan_et_al2012}, and \citealt{crossfield_et_al2013} for
further discussion).  All one-dimensional models of sufficiently
highly irradiated planets have deep, nearly isothermal, layers,
extending more than two decades in pressure down from the photosphere
at $\sim$1~bar.  In general, the temperature of the isothermal layer
scales with the strength of the irradiation roughly in proportion to
$F_0^{1/4}$, where $F_0$ is the incident flux at the substellar point
\citep{hubeny_et_al2003}.  However, the presence of a thermal
inversion tends to heat the upper atmosphere and cool the lower
atmosphere relative to the profile that would obtain without an
inversion, as is clear from examining the relative temperature
profiles of, e.g., WASP-18b and WASP-12b.  The electron mixing ratio
is calculated via Saha equations, assuming solar composition of the
elements. The conductivity is calculated with
equation~(\ref{eq:conductivity}), found in \citet{menou2012}.

In order to produce the heating profile plot in the lower right panel
of Fig.~\ref{fig:dayprof}, we use equation~(\ref{eq:PowerDensity}) to
calculate the ohmic heating density, and vertically integrate this
density times $\pi R_p^2$ from the outside of the planet inward.  The
abscissa, then, is the ratio of the cumulative ohmic power (from the
outside of the planet down to a given ordinate) to the incident
stellar power.  Note that the ratio of the full integral of the ohmic
power to the incident irradiation is the efficiency factor
$\varepsilon$ referred to in \S\ref{ssec:stable} below.  The
temperature and electron mixing ratio profiles are calculated from the
model atmospheres, and the zonal wind speed $u$ and magnetic field
strength $B$ are fixed at values motivated by previous studies in the
literature.  We present two heating profiles for each planet.  The
solid curve assumes zonal wind speeds of 1~km~s$^{-1}$ throughout the
whole atmosphere, where this speed is motivated by the results of a
variety of circulation models in the literature and by the simple
scalings in equations~(\ref{eq:MomEq}) and (\ref{eq:rewrite}).  The
dashed curve assumes that the zonal wind speed is prograde (relative
to the co-rotating frame) by 1~km~s$^{-1}$ at pressures lower than
$10^{-4}$~bars, stationary (0~km~s$^{-1}$) at pressures greater than
300~bars, and varies linearly with altitude in between.  The magnetic
fields are assumed to be 3~G for all planets.  Note that increasing
$B$ from 3~G to 10~G increases the heating by roughly an order of
magnitude, which results in the cumulative ohmic power being greater
than the incident stellar power for the models of HAT-P-7b and
WASP-12b if the winds are $\sim$1~km~s$^{-1}$ over a significant
portion of the day-side atmosphere.  In truth, the wind speeds would
depend on the magnetic field strength, which itself depends on
hydrodynamic motions in the deep interior and in the atmosphere that,
presumably, lead to a magnetic dynamo.  Numerous works have attempted
to simulate or calculate aspects of this feedback, with varying levels
of sophistication, although none of these works captures the full
process \citep{perna_et_al2010a, perna_et_al2010b, batygin_et_al2011,
  menou2012, rauscher+menou2013, wu+lithwick2013}.  The profiles
presented here should be taken not as quantitative predictions, but
rather as qualitative guides to the vertical variation in heating that
might obtain in hot Jupiter atmospheres at a range of levels of
irradiation.  These profiles may be scaled with $u^2 B^2$ for
different assumptions of wind speeds and magnetic field strengths.
Clearly, the largest contribution to the degree of ohmic heating is
the local conductivity, which varies by $\sim$8 orders of magnitude
across different altitudes and different planets.

The ionization depends very sensitively on temperature.  On the
nightsides of these planets (calculated, but not shown), the
temperatures are cool enough that electron mixing ratios and,
therefore, electrical conductivities, are much lower than on the
daysides.  The night-side conductivities are all
$\lsim$$10^4$~s$^{-1}$, rendering any implied heating rate too low
($<$$10^{-4}$ of the incident stellar power) to have a significant
effect on planet radius.

Our one-dimensional atmosphere models of the day and night sides of
highly irradiated gas-giant planets suggest that the nightsides might
have very low free-electron fractions and experience negligible ohmic
heating.  The daysides of the hottest hot Jupiters, however, should be
highly enough ionized that that they might experience significant
ohmic heating if wind speeds are as high as dynamical models have
suggested and if the magnetic field strengths are as great as
Jupiter's $\sim$4-Gauss field \citep{russell1993,
  christensen_et_al2009}.

\subsection{Planetary Radii Are Generally Stable Against Runaway
  Expansion Due to Ohmic Heating}
\label{ssec:stable}

In order to investigate the stability of a planet's radius, we
consider how an equilibrium situation responds to a perturbation.
Define $R_p^0$ as the equilibrium radius.  That is, at $R_p = R_p^0$,
heating balances cooling and $dR_p/dt = 0$.  Let $s$ be the
dimensionless specific entropy per baryon --- in a fully convective,
approximately isentropic object, essentially the total entropy
$\mathcal{S}$ divided by the product of Boltzmann's constant $k_B$
with the total number of baryons $M_p/m_p$: $s = \mathcal{S}/(k_B
M_p/m_p)$.

Note that, in a differential mass element $dm$, a change in heat
${\dbar}q$ corresponds to a change in entropy via ${\dbar}q = k_B T
(d\mathcal{S}/dt)dt = k_B N_A T (ds/dt) dm dt$, where the operator
$\dbar$ indicates an incomplete differential and $N_A$ is Avogadro's
number.  So, the mass-integrated change in heat ${\dbar}Q$ may be
expressed as:
\begin{equation}
{\dbar}Q = \brp{\int_{\rm Mass} k_B N_A T \frac{ds}{dt}
  dm} dt = L_{\rm net-in} dt = -Ldt \, ,
\end{equation}
where $L_{\rm net-in} = -L$ is the net inward (heating) power, which
is the negative of $L$, the net outward (cooling) luminosity.  The net
luminosity $L = C - H$, where $C$ is the outward luminosity and $H$ is
the inward luminosity (and, at $R_p^0$, $C[R_p^0] = H[R_p^0]$).

These are the ingredients needed to evaluate $dR_p/dt$ --- i.e.,
$(dR_p/ds)(ds/dt)$ --- for a given perturbation $\Delta R_p$ from
$R_p^0$.  If we define $\delta_{R_p} \equiv \Delta R_p / R_p^0$, we find
that, unsurprisingly,
\be
\dot{\delta}_{R_p} \equiv \frac{d\delta_{R_p}}{dt} \sim - \frac{\delta_{R_p}}{\tau} \, ,
\ee
where the relaxation timescale $\tau$ may be written
\be
\tau  = \frac{\frac{1}{m_p}\int k_B T dm}{ \frac{dR}{ds} \frac{dL}{dR_p} }  \sim \frac{\widetilde{E}}{L} \, .
\ee
Here, $\widetilde{E}$ is of order the product of the number of
particles ($\sim$$M_p/m_p$) with the mass-weighted average of $k_B T$
($\sim$$M_p^{-1}\int k_B T dm$).  The form of $\tau$ makes clear that
it is essentially the Kelvin-Helmholtz timescale, which is the natural
timescale on which an object responds to thermal perturbations.
Importantly, a perturbation relaxes to equilibrium since the
coefficient of $\delta_{R_p}$ in the relation $\dot{\delta}_{R_p}
\propto \delta_{R_p}$ is negative so long as $dL/dR_p$ is positive.

\begin{figure*}[t]
\plotonesc{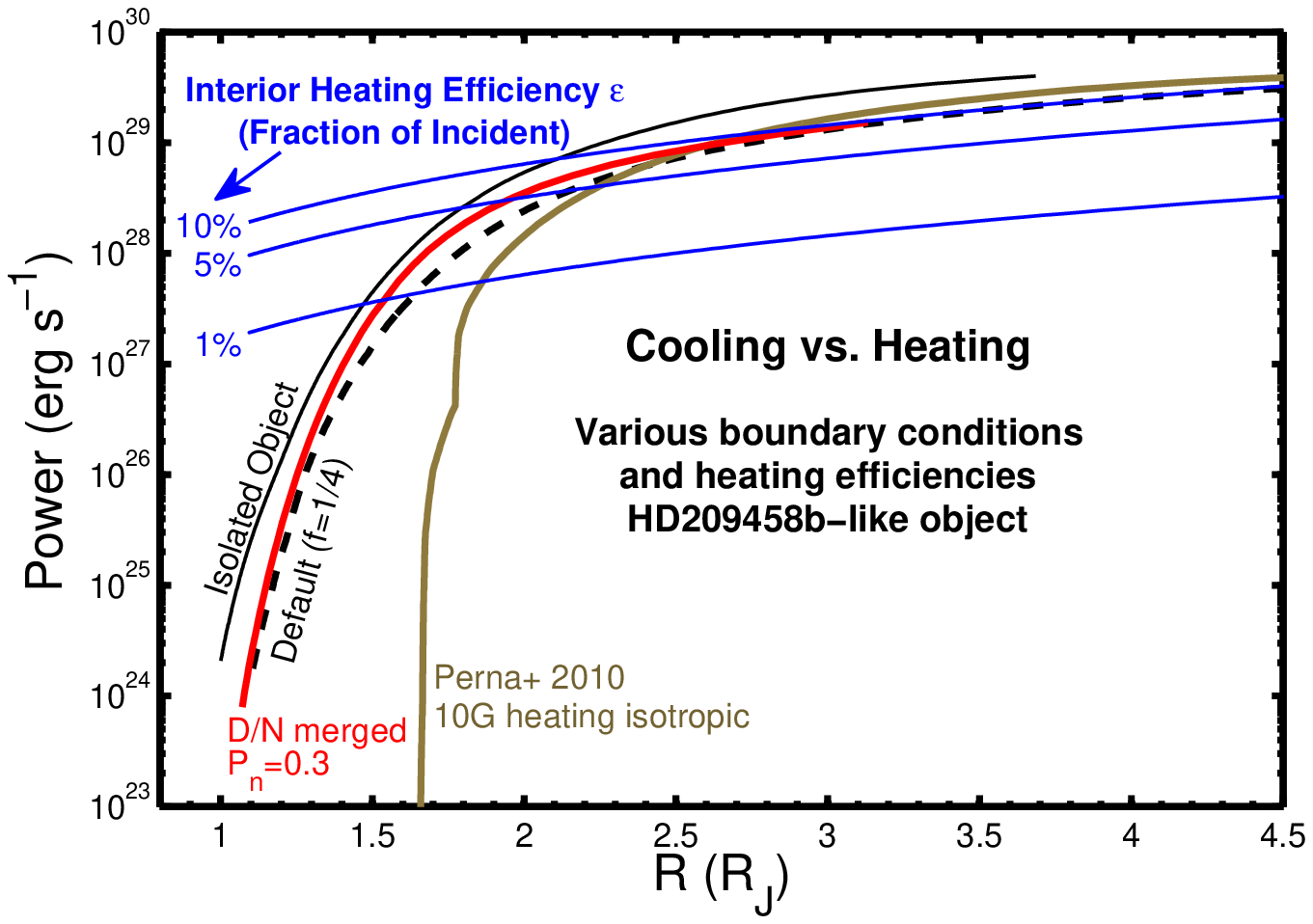}
\caption{Heating and cooling rates vs. total radius for irradiated
  models of an HD~209458b-like planet.  The heating curves (blue) show
  the heating power as a function of planetary radius, for three
  different assumed (constant) efficiencies of converting intercepted
  power to interior heating: 1\%, 5\%, and 10\%.  The cooling curves
  represent $4\pi R_p^2 \sigma T_{\rm eff}^4$, for four different
  cases: an isolated object (thin black) cooling with the boundary
  condition of \citet{burrows_et_al1997}; the 1-d $f=1/4$ boundary
  condition (dashed black) presented in Figs.~\ref{fig:kappa_effect}
  and \ref{fig:night_effect}--\ref{fig:heat_int_DN}; the 1+1-D
  boundary condition with $P_n=0.3$ (red) presented in
  Figs.~\ref{fig:night_effect} and \ref{fig:heat_int_atm_effect}; and
  a 1-D boundary condition assuming an atmospheric heating profile
  corresponding to 10~Gauss with drag from \citet{perna_et_al2010b}.
  For efficiencies less than $\sim$10\%, all cooling curves cross a
  heating curve at some (radius,power) pair, meaning that there is no
  purely thermal runaway instability in radius because the cooling
  power at some point balances (or exceeds) the heating power,
  although Roche-lobe overflow might occur when the radius becomes
  large enough, for planets that are particularly close-in.}
\label{fig:heatcool}
\end{figure*}

Figure~\ref{fig:heatcool} shows illustrative curves for heating power
versus cooling power for models of an irradiated hot Jupiter planet
(similar to HD~209458b).  Intersections between the heating curves and
the cooling curves represent equilibrium radii (``$R_P^0$'') where the
cooling balances the heating.  The upshot of the stability argument
just presented is evident in this figure, because at all equilibria
the slope of the cooling curve is greater than the slope of the
heating curve, indicating that a small positive perturbation to the
radius will lead to cooling exceeding heating and, therefore, to the
radius shrinking back to the equilibrium.  This indicates that, in
these models, there is no purely thermal runaway instability in radius
(for $\varepsilon \lsim 10\%$).  This result is in contrast to the
suggestion in \citet{batygin_et_al2011} that ohmic heating can lead to
a thermal instability that culminates in complete evaporation of a
planet.  When a planet's radius becomes large enough, however, a
runaway Roche-lobe overflow process can still occur that might
initially be triggered by ohmic heating.

\section{Conclusions}
\label{sec:conc}

We have presented an analysis of various subtle, but important,
physical effects that can influence the radii of hot Jupiters.  To
explain the anomalously large radii of an interesting subset of this
class, some have invoked the deposition of thermal power either in
planets' deep, convective interiors or at various levels in their
radiative atmospheres. To explore the consequences of such heating on
the radii and radius evolution of hot Jupiters, we generated
self-consistent atmosphere-planet evolutionary models and found that
the evolution of highly irradiated gas-giant planets depends
sensitively on where the energy is deposited.  If ``extra'' power goes
into heating the upper atmosphere, via an extra
thermal-inversion-causing optical absorber, this (slightly) increases
the evolutionary cooling rate and results in slightly smaller radii.
If the additional power is deposited more deeply in the radiative
atmosphere (e.g., via ohmic heating), the interior cooling rate
decreases, with the result that the planet's radius is larger at a
given age.  The radius-boosting effect of a given amount of power
increases with the depth at which that power is delivered in the
atmosphere. However, the asymptotic radius in the limit of infinite
age is not changed.  In other words, atmospheric heating might be able
to produce a large inflating effect, but one that does not last
forever.  In contrast, the same power deposited deep in the convective
interior of the planet always has a larger asymptotic radius than if
the same power is delivered anywhere in the radiative zone.

Importantly, cooling through the (cooler) nightside of a tidally
locked hot Jupiter, or through its (cooler) poles, increases the rate
of interior cooling and radius shrinkage relative to isotropic models.
As a result, the radius-inflating influence of atmospheric heating is
reduced in a 1+1-D model that consistently treats day/night
redistribution of heat and cooling through the nightside. This makes
it difficult to significantly increase the radius of a hot Jupiter via
atmospheric heating alone.

In addition, we find a limit on the product of zonal wind speed and
magnetic field strength in hot Jupiter atmospheres that have
significant ionization.  Specifically, the most highly irradiated
planets cannot have $uB \gtrsim 10 {\rm~km~s^{-1}~Gauss}$ over a large
fraction of their daysides, where $u$ is the zonal wind speed and $B$
is the dipolar magnetic field strength in the atmosphere.

The vexing problem of the inflated radii of some hot Jupiters has
persisted for more than a decade.  It has long been appreciated that
intense irradiation can slow the evolutionary cooling of these
objects, whatever other processes are invoked to explain the anomaly.
Here, however, we have shown that coupling the day-side cooling with
that through the nightside or the polar regions largely undoes the
reduced interior cooling effect of the irradiating flux.  An extra
power source may, then, be needed in their deep interiors, not merely
in their atmospheres, to explain most inflated, multi-billion-year-old
planets. Moreover, the degree of extra power that must be deposited in
the deep interior would seem to be greater than has been inferred in
those many studies using isotropically cooling models.

\acknowledgements

We thank Ivan Hubeny, Kristen Menou, Adam Showman, Jano Budaj, Yanqin
Wu, and Greg Novak for useful discussions.  The authors acknowledge
support in part under NASA ATP grant NNX07AG80G, HST grants
HST-GO-12181.04-A and HST-GO-12314.03-A, and JPL/Spitzer Agreements
1417122, 1348668, 1371432, and 1377197.  DSS gratefully acknowledges
support from NSF grant AST-0807444 and the Keck Fellowship, and the
Friends of the Institute.

\bibliography{biblio.bib}

\begin{thebibliography}{116}
\expandafter\ifx\csname natexlab\endcsname\relax\def\natexlab#1{#1}\fi

\bibitem[{{Anderson} {et~al.}(2011){Anderson}, {Smith}, {Lanotte}, {Barman},
  {Collier Cameron}, {Campo}, {Gillon}, {Harrington}, {Hellier}, {Maxted},
  {Queloz}, {Triaud}, \& {Wheatley}}]{anderson_et_al2011b}
{Anderson}, D.~R., {Smith}, A.~M.~S., {Lanotte}, A.~A., {Barman}, T.~S.,
  {Collier Cameron}, A., {Campo}, C.~J., {Gillon}, M., {Harrington}, J.,
  {Hellier}, C., {Maxted}, P.~F.~L., {Queloz}, D., {Triaud}, A.~H.~M.~J., \&
  {Wheatley}, P.~J. 2011, \mnras, 416, 2108

\bibitem[{{Arras} \& {Socrates}(2009{\natexlab{a}})}]{arras+socrates2009b}
{Arras}, P. \& {Socrates}, A. 2009{\natexlab{a}}, ArXiv e-prints
  arXiv:0912.2318

\bibitem[{{Arras} \& {Socrates}(2009{\natexlab{b}})}]{arras+socrates2009a}
---. 2009{\natexlab{b}}, ArXiv e-prints arXiv:0901.0735

\bibitem[{{Arras} \& {Socrates}(2010)}]{arras+socrates2010}
---. 2010, \apj, 714, 1

\bibitem[{{Baraffe} {et~al.}(2006){Baraffe}, {Alibert}, {Chabrier}, \&
  {Benz}}]{baraffe_et_al2006}
{Baraffe}, I., {Alibert}, Y., {Chabrier}, G., \& {Benz}, W. 2006, \aap, 450,
  1221

\bibitem[{{Baraffe} {et~al.}(2008){Baraffe}, {Chabrier}, \&
  {Barman}}]{baraffe_et_al2008}
{Baraffe}, I., {Chabrier}, G., \& {Barman}, T. 2008, \aap, 482, 315

\bibitem[{{Baraffe} {et~al.}(2010){Baraffe}, {Chabrier}, \&
  {Barman}}]{baraffe_et_al2010}
---. 2010, Reports on Progress in Physics, 73, 016901

\bibitem[{{Baraffe} {et~al.}(2003){Baraffe}, {Chabrier}, {Barman}, {Allard}, \&
  {Hauschildt}}]{baraffe_et_al2003}
{Baraffe}, I., {Chabrier}, G., {Barman}, T.~S., {Allard}, F., \& {Hauschildt},
  P.~H. 2003, \aap, 402, 701

\bibitem[{{Baraffe} {et~al.}(2005){Baraffe}, {Chabrier}, {Barman}, {Selsis},
  {Allard}, \& {Hauschildt}}]{baraffe_et_al2005}
{Baraffe}, I., {Chabrier}, G., {Barman}, T.~S., {Selsis}, F., {Allard}, F., \&
  {Hauschildt}, P.~H. 2005, \aap, 436, L47

\bibitem[{{Baraffe} {et~al.}(2004){Baraffe}, {Selsis}, {Chabrier}, {Barman},
  {Allard}, {Hauschildt}, \& {Lammer}}]{baraffe_et_al2004}
{Baraffe}, I., {Selsis}, F., {Chabrier}, G., {Barman}, T.~S., {Allard}, F.,
  {Hauschildt}, P.~H., \& {Lammer}, H. 2004, \aap, 419, L13

\bibitem[{{Batygin} \& {Stevenson}(2010)}]{batygin+stevenson2010}
{Batygin}, K. \& {Stevenson}, D.~J. 2010, \apjl, 714, L238

\bibitem[{{Batygin} {et~al.}(2011){Batygin}, {Stevenson}, \&
  {Bodenheimer}}]{batygin_et_al2011}
{Batygin}, K., {Stevenson}, D.~J., \& {Bodenheimer}, P.~H. 2011, \apj, 738, 1

\bibitem[{{Bodenheimer} {et~al.}(2003){Bodenheimer}, {Laughlin}, \&
  {Lin}}]{bodenheimer_et_al2003}
{Bodenheimer}, P., {Laughlin}, G., \& {Lin}, D.~N.~C. 2003, \apj, 592, 555

\bibitem[{{Bodenheimer} {et~al.}(2001){Bodenheimer}, {Lin}, \&
  {Mardling}}]{bodenheimer_et_al2001}
{Bodenheimer}, P., {Lin}, D.~N.~C., \& {Mardling}, R.~A. 2001, \apj, 548, 466

\bibitem[{{Boss}(1995)}]{boss1995}
{Boss}, A.~P. 1995, Science, 267, 360

\bibitem[{{Budaj}(2011)}]{Budaj2011}
{Budaj}, J. 2011, \aj, 141, 59

\bibitem[{{Budaj} {et~al.}(2012){Budaj}, {Hubeny}, \&
  {Burrows}}]{budaj_et_al2012}
{Budaj}, J., {Hubeny}, I., \& {Burrows}, A. 2012, \aap, 537, A115

\bibitem[{{Burrows} {et~al.}(2008){Burrows}, {Budaj}, \&
  {Hubeny}}]{burrows_et_al2008b}
{Burrows}, A., {Budaj}, J., \& {Hubeny}, I. 2008, \apj, 678, 1436

\bibitem[{{Burrows} {et~al.}(2000){Burrows}, {Guillot}, {Hubbard}, {Marley},
  {Saumon}, {Lunine}, \& {Sudarsky}}]{burrows_et_al2000}
{Burrows}, A., {Guillot}, T., {Hubbard}, W.~B., {Marley}, M.~S., {Saumon}, D.,
  {Lunine}, J.~I., \& {Sudarsky}, D. 2000, \apjl, 534, L97

\bibitem[{{Burrows} {et~al.}(2011){Burrows}, {Heng}, \&
  {Nampaisarn}}]{burrows_et_al2011}
{Burrows}, A., {Heng}, K., \& {Nampaisarn}, T. 2011, \apj, 736, 47

\bibitem[{{Burrows} {et~al.}(1994){Burrows}, {Hubbard}, \&
  {Lunine}}]{burrows_et_al1994}
{Burrows}, A., {Hubbard}, W.~B., \& {Lunine}, J.~I. 1994, in Astronomical
  Society of the Pacific Conference Series, Vol.~64, Cool Stars, Stellar
  Systems, and the Sun, ed. {J.-P.~Caillault}, 528--+

\bibitem[{{Burrows} {et~al.}(2001){Burrows}, {Hubbard}, {Lunine}, \&
  {Liebert}}]{burrows_et_al2001}
{Burrows}, A., {Hubbard}, W.~B., {Lunine}, J.~I., \& {Liebert}, J. 2001,
  Reviews of Modern Physics, 73, 719

\bibitem[{{Burrows} {et~al.}(2007{\natexlab{a}}){Burrows}, {Hubeny}, {Budaj},
  \& {Hubbard}}]{burrows_et_al2007}
{Burrows}, A., {Hubeny}, I., {Budaj}, J., \& {Hubbard}, W.~B.
  2007{\natexlab{a}}, \apj, 661, 502

\bibitem[{{Burrows} {et~al.}(2007{\natexlab{b}}){Burrows}, {Hubeny}, {Budaj},
  {Knutson}, \& {Charbonneau}}]{burrows_et_al2007c}
{Burrows}, A., {Hubeny}, I., {Budaj}, J., {Knutson}, H.~A., \& {Charbonneau},
  D. 2007{\natexlab{b}}, \apjl, 668, L171

\bibitem[{{Burrows} {et~al.}(2004){Burrows}, {Hubeny}, {Hubbard}, {Sudarsky},
  \& {Fortney}}]{burrows_et_al2004}
{Burrows}, A., {Hubeny}, I., {Hubbard}, W.~B., {Sudarsky}, D., \& {Fortney},
  J.~J. 2004, \apjl, 610, L53

\bibitem[{{Burrows} \& {Liebert}(1993)}]{burrows+liebert1993}
{Burrows}, A. \& {Liebert}, J. 1993, Reviews of Modern Physics, 65, 301

\bibitem[{{Burrows} {et~al.}(1997){Burrows}, {Marley}, {Hubbard}, {Lunine},
  {Guillot}, {Saumon}, {Freedman}, {Sudarsky}, \& {Sharp}}]{burrows_et_al1997}
{Burrows}, A., {Marley}, M., {Hubbard}, W.~B., {Lunine}, J.~I., {Guillot}, T.,
  {Saumon}, D., {Freedman}, R., {Sudarsky}, D., \& {Sharp}, C. 1997, \apj, 491,
  856

\bibitem[{{Burrows} {et~al.}(2010){Burrows}, {Rauscher}, {Spiegel}, \&
  {Menou}}]{burrows_et_al2010}
{Burrows}, A., {Rauscher}, E., {Spiegel}, D.~S., \& {Menou}, K. 2010, \apj,
  719, 341

\bibitem[{{Burrows} {et~al.}(2003){Burrows}, {Sudarsky}, \&
  {Hubbard}}]{burrows_et_al2003}
{Burrows}, A., {Sudarsky}, D., \& {Hubbard}, W.~B. 2003, \apj, 594, 545

\bibitem[{{Burrows} {et~al.}(2006){Burrows}, {Sudarsky}, \&
  {Hubeny}}]{burrows_et_al2006}
{Burrows}, A., {Sudarsky}, D., \& {Hubeny}, I. 2006, \apj, 650, 1140

\bibitem[{{Chabrier} \& {Baraffe}(2007)}]{chabrier+baraffe2007}
{Chabrier}, G. \& {Baraffe}, I. 2007, \apjl, 661, L81

\bibitem[{{Chabrier} {et~al.}(2004){Chabrier}, {Barman}, {Baraffe}, {Allard},
  \& {Hauschildt}}]{chabrier_et_al2004}
{Chabrier}, G., {Barman}, T., {Baraffe}, I., {Allard}, F., \& {Hauschildt},
  P.~H. 2004, \apjl, 603, L53

\bibitem[{{Chan} {et~al.}(2011){Chan}, {Ingemyr}, {Winn}, {Holman},
  {Sanchis-Ojeda}, {Esquerdo}, \& {Everett}}]{chan_et_al2011}
{Chan}, T., {Ingemyr}, M., {Winn}, J.~N., {Holman}, M.~J., {Sanchis-Ojeda}, R.,
  {Esquerdo}, G., \& {Everett}, M. 2011, \aj, 141, 179

\bibitem[{{Charbonneau} {et~al.}(2000){Charbonneau}, {Brown}, {Latham}, \&
  {Mayor}}]{charbonneau_et_al2000}
{Charbonneau}, D., {Brown}, T.~M., {Latham}, D.~W., \& {Mayor}, M. 2000, \apjl,
  529, L45

\bibitem[{{Christensen} {et~al.}(2009){Christensen}, {Holzwarth}, \&
  {Reiners}}]{christensen_et_al2009}
{Christensen}, U.~R., {Holzwarth}, V., \& {Reiners}, A. 2009, \nat, 457, 167

\bibitem[{{Cooper} \& {Showman}(2005)}]{cooper+showman2005}
{Cooper}, C.~S. \& {Showman}, A.~P. 2005, \apjl, 629, L45

\bibitem[{{Cowan} \& {Agol}(2011)}]{cowan+agol2011}
{Cowan}, N.~B. \& {Agol}, E. 2011, \apj, 729, 54

\bibitem[{{Cowan} {et~al.}(2012){Cowan}, {Machalek}, {Croll}, {Shekhtman},
  {Burrows}, {Deming}, {Greene}, \& {Hora}}]{cowan_et_al2012}
{Cowan}, N.~B., {Machalek}, P., {Croll}, B., {Shekhtman}, L.~M., {Burrows}, A.,
  {Deming}, D., {Greene}, T., \& {Hora}, J.~L. 2012, \apj, 747, 82

\bibitem[{{Crossfield} {et~al.}(2013){Crossfield}, {Barman}, {Hansen},
  {Tanaka}, \& {Kodama}}]{crossfield_et_al2013}
{Crossfield}, I.~J.~M., {Barman}, T., {Hansen}, B.~M.~S., {Tanaka}, I., \&
  {Kodama}, T. 2013, ArXiv e-prints

\bibitem[{{Demory} \& {Seager}(2011)}]{demory+seager2011}
{Demory}, B.-O. \& {Seager}, S. 2011, \apjs, 197, 12

\bibitem[{{Dobbs-Dixon} \& {Lin}(2008)}]{dobbs-dixon+lin2008}
{Dobbs-Dixon}, I. \& {Lin}, D.~N.~C. 2008, \apj, 673, 513

\bibitem[{{Draine} {et~al.}(1983){Draine}, {Roberge}, \&
  {Dalgarno}}]{draine_et_al1983}
{Draine}, B.~T., {Roberge}, W.~G., \& {Dalgarno}, A. 1983, \apj, 264, 485

\bibitem[{{Fortney} \& {Hubbard}(2004)}]{fortney+hubbard2004}
{Fortney}, J.~J. \& {Hubbard}, W.~B. 2004, \apj, 608, 1039

\bibitem[{{Fortney} {et~al.}(2011){Fortney}, {Ikoma}, {Nettelmann}, {Guillot},
  \& {Marley}}]{fortney_et_al2011}
{Fortney}, J.~J., {Ikoma}, M., {Nettelmann}, N., {Guillot}, T., \& {Marley},
  M.~S. 2011, \apj, 729, 32

\bibitem[{{Fortney} {et~al.}(2008{\natexlab{a}}){Fortney}, {Lodders}, {Marley},
  \& {Freedman}}]{fortney_et_al2008}
{Fortney}, J.~J., {Lodders}, K., {Marley}, M.~S., \& {Freedman}, R.~S.
  2008{\natexlab{a}}, \apj, 678, 1419

\bibitem[{{Fortney} \& {Marley}(2007)}]{fortney_et_al2007}
{Fortney}, J.~J. \& {Marley}, M.~S. 2007, \apjl, 666, L45

\bibitem[{{Fortney} {et~al.}(2008{\natexlab{b}}){Fortney}, {Marley}, {Saumon},
  \& {Lodders}}]{fortney_et_al2008b}
{Fortney}, J.~J., {Marley}, M.~S., {Saumon}, D., \& {Lodders}, K.
  2008{\natexlab{b}}, \apj, 683, 1104

\bibitem[{{Fressin} {et~al.}(2010){Fressin}, {Knutson}, {Charbonneau},
  {O'Donovan}, {Burrows}, {Deming}, {Mandushev}, \&
  {Spiegel}}]{fressin_et_al2010}
{Fressin}, F., {Knutson}, H.~A., {Charbonneau}, D., {O'Donovan}, F.~T.,
  {Burrows}, A., {Deming}, D., {Mandushev}, G., \& {Spiegel}, D. 2010, \apj,
  711, 374

\bibitem[{{Goodman} \& {Lackner}(2009)}]{goodman+lackner2009}
{Goodman}, J. \& {Lackner}, C. 2009, \apj, 696, 2054

\bibitem[{{Gu} {et~al.}(2003){Gu}, {Lin}, \& {Bodenheimer}}]{gu_et_al2003}
{Gu}, P.-G., {Lin}, D.~N.~C., \& {Bodenheimer}, P.~H. 2003, \apj, 588, 509

\bibitem[{{Guillot}(2010)}]{guillot2010}
{Guillot}, T. 2010, \aap, 520, A27+

\bibitem[{{Guillot} {et~al.}(1996){Guillot}, {Burrows}, {Hubbard}, {Lunine}, \&
  {Saumon}}]{guillot_et_al1996}
{Guillot}, T., {Burrows}, A., {Hubbard}, W.~B., {Lunine}, J.~I., \& {Saumon},
  D. 1996, \apjl, 459, L35+

\bibitem[{{Guillot} {et~al.}(2006){Guillot}, {Santos}, {Pont}, {Iro}, {Melo},
  \& {Ribas}}]{guillot2006}
{Guillot}, T., {Santos}, N.~C., {Pont}, F., {Iro}, N., {Melo}, C., \& {Ribas},
  I. 2006, \aap, 453, L21

\bibitem[{{Guillot} \& {Showman}(2002)}]{guillot+showman2002}
{Guillot}, T. \& {Showman}, A.~P. 2002, \aap, 385, 156

\bibitem[{{Hansen}(2008)}]{hansen2008}
{Hansen}, B.~M.~S. 2008, \apjs, 179, 484

\bibitem[{{Hartman} {et~al.}(2011){Hartman}, {Bakos}, {Torres}, {Latham},
  {Kov{\'a}cs}, {B{\'e}ky}, {Quinn}, {Mazeh}, {Shporer}, {Marcy}, {Howard},
  {Fischer}, {Johnson}, {Esquerdo}, {Noyes}, {Sasselov}, {Stefanik},
  {Fernandez}, {Szklen{\'a}r}, {L{\'a}z{\'a}r}, {Papp}, \&
  {S{\'a}ri}}]{hartman_et_al2011}
{Hartman}, J.~D., {Bakos}, G.~{\'A}., {Torres}, G., {Latham}, D.~W.,
  {Kov{\'a}cs}, G., {B{\'e}ky}, B., {Quinn}, S.~N., {Mazeh}, T., {Shporer}, A.,
  {Marcy}, G.~W., {Howard}, A.~W., {Fischer}, D.~A., {Johnson}, J.~A.,
  {Esquerdo}, G.~A., {Noyes}, R.~W., {Sasselov}, D.~D., {Stefanik}, R.~P.,
  {Fernandez}, J.~M., {Szklen{\'a}r}, T., {L{\'a}z{\'a}r}, J., {Papp}, I., \&
  {S{\'a}ri}, P. 2011, \apj, 742, 59

\bibitem[{{Hebb} {et~al.}(2009){Hebb}, {Collier-Cameron}, {Loeillet},
  {Pollacco}, {H{\'e}brard}, {Street}, {Bouchy}, {Stempels}, {Moutou},
  {Simpson}, {Udry}, {Joshi}, {West}, {Skillen}, {Wilson}, {McDonald},
  {Gibson}, {Aigrain}, {Anderson}, {Benn}, {Christian}, {Enoch}, {Haswell},
  {Hellier}, {Horne}, {Irwin}, {Lister}, {Maxted}, {Mayor}, {Norton}, {Parley},
  {Pont}, {Queloz}, {Smalley}, \& {Wheatley}}]{hebb_et_al2009}
{Hebb}, L., {Collier-Cameron}, A., {Loeillet}, B., {Pollacco}, D.,
  {H{\'e}brard}, G., {Street}, R.~A., {Bouchy}, F., {Stempels}, H.~C.,
  {Moutou}, C., {Simpson}, E., {Udry}, S., {Joshi}, Y.~C., {West}, R.~G.,
  {Skillen}, I., {Wilson}, D.~M., {McDonald}, I., {Gibson}, N.~P., {Aigrain},
  S., {Anderson}, D.~R., {Benn}, C.~R., {Christian}, D.~J., {Enoch}, B.,
  {Haswell}, C.~A., {Hellier}, C., {Horne}, K., {Irwin}, J., {Lister}, T.~A.,
  {Maxted}, P., {Mayor}, M., {Norton}, A.~J., {Parley}, N., {Pont}, F.,
  {Queloz}, D., {Smalley}, B., \& {Wheatley}, P.~J. 2009, \apj, 693, 1920

\bibitem[{{Heng} {et~al.}(2011){Heng}, {Menou}, \&
  {Phillipps}}]{heng_et_al2011}
{Heng}, K., {Menou}, K., \& {Phillipps}, P.~J. 2011, \mnras, 413, 2380

\bibitem[{{Henry} {et~al.}(2000){Henry}, {Marcy}, {Butler}, \&
  {Vogt}}]{henry_et_al2000}
{Henry}, G.~W., {Marcy}, G.~W., {Butler}, R.~P., \& {Vogt}, S.~S. 2000, \apjl,
  529, L41

\bibitem[{{Howard} {et~al.}(2012){Howard}, {Marcy}, {Bryson}, {Jenkins},
  {Rowe}, {Batalha}, {Borucki}, {Koch}, {Dunham}, {Gautier}, {Van Cleve},
  {Cochran}, {Latham}, {Lissauer}, {Torres}, {Brown}, {Gilliland}, {Buchhave},
  {Caldwell}, {Christensen-Dalsgaard}, {Ciardi}, {Fressin}, {Haas}, {Howell},
  {Kjeldsen}, {Seager}, {Rogers}, {Sasselov}, {Steffen}, {Basri},
  {Charbonneau}, {Christiansen}, {Clarke}, {Dupree}, {Fabrycky}, {Fischer},
  {Ford}, {Fortney}, {Tarter}, {Girouard}, {Holman}, {Johnson}, {Klaus},
  {Machalek}, {Moorhead}, {Morehead}, {Ragozzine}, {Tenenbaum}, {Twicken},
  {Quinn}, {Isaacson}, {Shporer}, {Lucas}, {Walkowicz}, {Welsh}, {Boss},
  {Devore}, {Gould}, {Smith}, {Morris}, {Prsa}, {Morton}, {Still}, {Thompson},
  {Mullally}, {Endl}, \& {MacQueen}}]{howard_et_al2012}
{Howard}, A.~W., {Marcy}, G.~W., {Bryson}, S.~T., {Jenkins}, J.~M., {Rowe},
  J.~F., {Batalha}, N.~M., {Borucki}, W.~J., {Koch}, D.~G., {Dunham}, E.~W.,
  {Gautier}, III, T.~N., {Van Cleve}, J., {Cochran}, W.~D., {Latham}, D.~W.,
  {Lissauer}, J.~J., {Torres}, G., {Brown}, T.~M., {Gilliland}, R.~L.,
  {Buchhave}, L.~A., {Caldwell}, D.~A., {Christensen-Dalsgaard}, J., {Ciardi},
  D., {Fressin}, F., {Haas}, M.~R., {Howell}, S.~B., {Kjeldsen}, H., {Seager},
  S., {Rogers}, L., {Sasselov}, D.~D., {Steffen}, J.~H., {Basri}, G.~S.,
  {Charbonneau}, D., {Christiansen}, J., {Clarke}, B., {Dupree}, A.,
  {Fabrycky}, D.~C., {Fischer}, D.~A., {Ford}, E.~B., {Fortney}, J.~J.,
  {Tarter}, J., {Girouard}, F.~R., {Holman}, M.~J., {Johnson}, J.~A., {Klaus},
  T.~C., {Machalek}, P., {Moorhead}, A.~V., {Morehead}, R.~C., {Ragozzine}, D.,
  {Tenenbaum}, P., {Twicken}, J.~D., {Quinn}, S.~N., {Isaacson}, H., {Shporer},
  A., {Lucas}, P.~W., {Walkowicz}, L.~M., {Welsh}, W.~F., {Boss}, A., {Devore},
  E., {Gould}, A., {Smith}, J.~C., {Morris}, R.~L., {Prsa}, A., {Morton},
  T.~D., {Still}, M., {Thompson}, S.~E., {Mullally}, F., {Endl}, M., \&
  {MacQueen}, P.~J. 2012, \apjs, 201, 15

\bibitem[{{Huang} \& {Cumming}(2012)}]{huang+cumming2012}
{Huang}, X. \& {Cumming}, A. 2012, \apj, 757, 47

\bibitem[{{Hubeny}(1988)}]{hubeny1988}
{Hubeny}, I. 1988, Computer Physics Communications, 52, 103

\bibitem[{{Hubeny} {et~al.}(2003){Hubeny}, {Burrows}, \&
  {Sudarsky}}]{hubeny_et_al2003}
{Hubeny}, I., {Burrows}, A., \& {Sudarsky}, D. 2003, \apj, 594, 1011

\bibitem[{{Hubeny} \& {Lanz}(1995)}]{hubeny+lanz1995}
{Hubeny}, I. \& {Lanz}, T. 1995, \apj, 439, 875

\bibitem[{{Ibgui} \& {Burrows}(2009)}]{ibgui+burrows2009}
{Ibgui}, L. \& {Burrows}, A. 2009, \apj, 700, 1921

\bibitem[{{Ibgui} {et~al.}(2010){Ibgui}, {Burrows}, \&
  {Spiegel}}]{ibgui_et_al2010}
{Ibgui}, L., {Burrows}, A., \& {Spiegel}, D.~S. 2010, \apj, 713, 751

\bibitem[{{Ibgui} {et~al.}(2011){Ibgui}, {Spiegel}, \&
  {Burrows}}]{ibgui_et_al2011}
{Ibgui}, L., {Spiegel}, D.~S., \& {Burrows}, A. 2011, \apj, 727, 75

\bibitem[{{Jackson} {et~al.}(2008){Jackson}, {Greenberg}, \&
  {Barnes}}]{jackson_et_al2008c}
{Jackson}, B., {Greenberg}, R., \& {Barnes}, R. 2008, \apj, 681, 1631

\bibitem[{{Knutson} {et~al.}(2008){Knutson}, {Charbonneau}, {Allen}, {Burrows},
  \& {Megeath}}]{knutson_et_al2008b}
{Knutson}, H.~A., {Charbonneau}, D., {Allen}, L.~E., {Burrows}, A., \&
  {Megeath}, S.~T. 2008, \apj, 673, 526

\bibitem[{{Knutson} {et~al.}(2010){Knutson}, {Howard}, \&
  {Isaacson}}]{knutson_et_al2010}
{Knutson}, H.~A., {Howard}, A.~W., \& {Isaacson}, H. 2010, \apj, 720, 1569

\bibitem[{{Langton} \& {Laughlin}(2008)}]{langton+laughlin2008}
{Langton}, J. \& {Laughlin}, G. 2008, \apj, 674, 1106

\bibitem[{{Laughlin} {et~al.}(2011){Laughlin}, {Crismani}, \&
  {Adams}}]{laughlin_et_al2011}
{Laughlin}, G., {Crismani}, M., \& {Adams}, F.~C. 2011, \apjl, 729, L7

\bibitem[{{Laughlin} {et~al.}(2005){Laughlin}, {Wolf}, {Vanmunster},
  {Bodenheimer}, {Fischer}, {Marcy}, {Butler}, \& {Vogt}}]{laughlin_et_al2005a}
{Laughlin}, G., {Wolf}, A., {Vanmunster}, T., {Bodenheimer}, P., {Fischer}, D.,
  {Marcy}, G., {Butler}, P., \& {Vogt}, S. 2005, \apj, 621, 1072

\bibitem[{{Leconte} \& {Chabrier}(2012)}]{leconte+chabrier2012}
{Leconte}, J. \& {Chabrier}, G. 2012, \aap, 540, A20

\bibitem[{{Leconte} {et~al.}(2010){Leconte}, {Chabrier}, {Baraffe}, \&
  {Levrard}}]{leconte_et_al2010}
{Leconte}, J., {Chabrier}, G., {Baraffe}, I., \& {Levrard}, B. 2010, \aap, 516,
  A64

\bibitem[{{Lissauer}(1995)}]{lissauer1995}
{Lissauer}, J.~J. 1995, Icarus, 114, 217

\bibitem[{{Liu} {et~al.}(2008){Liu}, {Burrows}, \& {Ibgui}}]{liu_et_al2008}
{Liu}, X., {Burrows}, A., \& {Ibgui}, L. 2008, \apj, 687, 1191

\bibitem[{{Lubow} {et~al.}(1997){Lubow}, {Tout}, \& {Livio}}]{lubow_et_al1997}
{Lubow}, S.~H., {Tout}, C.~A., \& {Livio}, M. 1997, \apj, 484, 866

\bibitem[{{Machalek} {et~al.}(2010 submitted){Machalek}, {Greene},
  {Harrington}, {Burrows}, \& {Hubeny}}]{machalek_et_al2010}
{Machalek}, P., {Greene}, T., {Harrington}, J., {Burrows}, A., \& {Hubeny}, I.
  2010 submitted, \apj

\bibitem[{{Madhusudhan}(2012)}]{madhusudhan2012}
{Madhusudhan}, N. 2012, \apj, 758, 36

\bibitem[{{Madhusudhan} {et~al.}(2011){Madhusudhan}, {Harrington}, {Stevenson},
  {Nymeyer}, {Campo}, {Wheatley}, {Deming}, {Blecic}, {Hardy}, {Lust},
  {Anderson}, {Collier-Cameron}, {Britt}, {Bowman}, {Hebb}, {Hellier},
  {Maxted}, {Pollacco}, \& {West}}]{madhusudhan_et_al2011b}
{Madhusudhan}, N., {Harrington}, J., {Stevenson}, K.~B., {Nymeyer}, S.,
  {Campo}, C.~J., {Wheatley}, P.~J., {Deming}, D., {Blecic}, J., {Hardy},
  R.~A., {Lust}, N.~B., {Anderson}, D.~R., {Collier-Cameron}, A., {Britt},
  C.~B.~T., {Bowman}, W.~C., {Hebb}, L., {Hellier}, C., {Maxted}, P.~F.~L.,
  {Pollacco}, D., \& {West}, R.~G. 2011, \nat, 469, 64

\bibitem[{{Madhusudhan} \& {Seager}(2009)}]{madhusudhan+seager2009}
{Madhusudhan}, N. \& {Seager}, S. 2009, \apj, 707, 24

\bibitem[{{Madhusudhan} \& {Seager}(2010)}]{madhusudhan+seager2010b}
---. 2010, \apj, 725, 261

\bibitem[{{Marcy} \& {Butler}(1996)}]{marcy+butler1996}
{Marcy}, G.~W. \& {Butler}, R.~P. 1996, \apjl, 464, L147+

\bibitem[{{Marleau} \& {Cumming}(2013)}]{marleau+cumming2013}
{Marleau}, G.-D. \& {Cumming}, A. 2013, ArXiv e-prints arXiv:1302.1517

\bibitem[{{Marley} {et~al.}(2007){Marley}, {Fortney}, {Hubickyj},
  {Bodenheimer}, \& {Lissauer}}]{marley_et_al2007}
{Marley}, M.~S., {Fortney}, J.~J., {Hubickyj}, O., {Bodenheimer}, P., \&
  {Lissauer}, J.~J. 2007, \apj, 655, 541

\bibitem[{{Mayor} \& {Queloz}(1995)}]{mayor+queloz1995}
{Mayor}, M. \& {Queloz}, D. 1995, \nat, 378, 355

\bibitem[{{Menou}(2012{\natexlab{a}})}]{menou2012}
{Menou}, K. 2012{\natexlab{a}}, \apj, 745, 138

\bibitem[{{Menou}(2012{\natexlab{b}})}]{menou2012b}
---. 2012{\natexlab{b}}, \apjl, 754, L9

\bibitem[{{Menou} \& {Rauscher}(2009)}]{menou+rauscher2009}
{Menou}, K. \& {Rauscher}, E. 2009, \apj, 700, 887

\bibitem[{{Miller} {et~al.}(2009){Miller}, {Fortney}, \&
  {Jackson}}]{miller_et_al2009}
{Miller}, N., {Fortney}, J.~J., \& {Jackson}, B. 2009, \apj, 702, 1413

\bibitem[{{Ogilvie} \& {Lin}(2007)}]{ogilvie+lin2007}
{Ogilvie}, G.~I. \& {Lin}, D.~N.~C. 2007, \apj, 661, 1180

\bibitem[{{Parmentier} {et~al.}(2013){Parmentier}, {Showman}, \&
  {Lian}}]{parmentier_et_al2013}
{Parmentier}, V., {Showman}, A.~P., \& {Lian}, Y. 2013, ArXiv e-prints
  arXiv:1301.4522

\bibitem[{{Paxton} {et~al.}(2013){Paxton}, {Cantiello}, {Arras}, {Bildsten},
  {Brown}, {Dotter}, {Mankovich}, {Montgomery}, {Stello}, {Timmes}, \&
  {Townsend}}]{paxton_et_al2013}
{Paxton}, B., {Cantiello}, M., {Arras}, P., {Bildsten}, L., {Brown}, E.~F.,
  {Dotter}, A., {Mankovich}, C., {Montgomery}, M.~H., {Stello}, D., {Timmes},
  F.~X., \& {Townsend}, R. 2013, ArXiv e-prints arXiv:1301.0319

\bibitem[{{Perna} {et~al.}(2010{\natexlab{a}}){Perna}, {Menou}, \&
  {Rauscher}}]{perna_et_al2010a}
{Perna}, R., {Menou}, K., \& {Rauscher}, E. 2010{\natexlab{a}}, \apj, 719, 1421

\bibitem[{{Perna} {et~al.}(2010{\natexlab{b}}){Perna}, {Menou}, \&
  {Rauscher}}]{perna_et_al2010b}
---. 2010{\natexlab{b}}, \apj, 724, 313

\bibitem[{{Rauscher} \& {Menou}(2013)}]{rauscher+menou2013}
{Rauscher}, E. \& {Menou}, K. 2013, \apj, 764, 103

\bibitem[{{Rein}(2012)}]{rein2012}
{Rein}, H. 2012, ArXiv e-prints arXiv:1211.7121

\bibitem[{{Russell}(1993)}]{russell1993}
{Russell}, C.~T. 1993, Reports on Progress in Physics, 56, 687

\bibitem[{{Schneider} {et~al.}(2011){Schneider}, {Dedieu}, {Le Sidaner},
  {Savalle}, \& {Zolotukhin}}]{schneider_et_al2011}
{Schneider}, J., {Dedieu}, C., {Le Sidaner}, P., {Savalle}, R., \&
  {Zolotukhin}, I. 2011, \aap, 532, A79

\bibitem[{{Sharp} \& {Burrows}(2007)}]{sharp+burrows2007}
{Sharp}, C.~M. \& {Burrows}, A. 2007, \apjs, 168, 140

\bibitem[{{Showman} {et~al.}(2011){Showman}, {Cho}, \&
  {Menou}}]{showman_et_al2011}
{Showman}, A.~P., {Cho}, J.~Y.-K., \& {Menou}, K. {Atmospheric Circulation of
  Exoplanets}, ed. S.~{Seager}, 471--516

\bibitem[{{Showman} {et~al.}(2009){Showman}, {Fortney}, {Lian}, {Marley},
  {Freedman}, {Knutson}, \& {Charbonneau}}]{showman_et_al2009}
{Showman}, A.~P., {Fortney}, J.~J., {Lian}, Y., {Marley}, M.~S., {Freedman},
  R.~S., {Knutson}, H.~A., \& {Charbonneau}, D. 2009, \apj, 699, 564

\bibitem[{{Showman} \& {Guillot}(2002)}]{showman+guillot2002}
{Showman}, A.~P. \& {Guillot}, T. 2002, \aap, 385, 166

\bibitem[{{Socrates}(2013)}]{socrates2013}
{Socrates}, A. 2013, ArXiv e-prints arXiv:1304.4121

\bibitem[{{Spiegel} \& {Burrows}(2010)}]{spiegel+burrows2010}
{Spiegel}, D.~S. \& {Burrows}, A. 2010, \apj, 722, 871

\bibitem[{{Spiegel} \& {Burrows}(2012)}]{spiegel+burrows2012}
---. 2012, \apj, 745, 174

\bibitem[{{Spiegel} {et~al.}(2011){Spiegel}, {Burrows}, \&
  {Milsom}}]{spiegel_et_al2011a}
{Spiegel}, D.~S., {Burrows}, A., \& {Milsom}, J.~A. 2011, \apj, 727, 57

\bibitem[{{Spiegel} \& {Madhusudhan}(2012)}]{spiegel+madhusudhan2012}
{Spiegel}, D.~S. \& {Madhusudhan}, N. 2012, \apj, 756, 132

\bibitem[{{Spiegel} {et~al.}(2009){Spiegel}, {Silverio}, \&
  {Burrows}}]{spiegel_et_al2009b}
{Spiegel}, D.~S., {Silverio}, K., \& {Burrows}, A. 2009, \apj, 699, 1487

\bibitem[{{Struve}(1952)}]{struve1952}
{Struve}, O. 1952, The Observatory, 72, 199

\bibitem[{{Wright} {et~al.}(2011){Wright}, {Fakhouri}, {Marcy}, {Han}, {Feng},
  {Johnson}, {Howard}, {Fischer}, {Valenti}, {Anderson}, \&
  {Piskunov}}]{wright_et_al2011}
{Wright}, J.~T., {Fakhouri}, O., {Marcy}, G.~W., {Han}, E., {Feng}, Y.,
  {Johnson}, J.~A., {Howard}, A.~W., {Fischer}, D.~A., {Valenti}, J.~A.,
  {Anderson}, J., \& {Piskunov}, N. 2011, \pasp, 123, 412

\bibitem[{{Wu}(2005{\natexlab{a}})}]{Wu_2005_1}
{Wu}, Y. 2005{\natexlab{a}}, \apj, 635, 674

\bibitem[{{Wu}(2005{\natexlab{b}})}]{Wu_2005_2}
---. 2005{\natexlab{b}}, \apj, 635, 688

\bibitem[{{Wu} \& {Lithwick}(2013)}]{wu+lithwick2013}
{Wu}, Y. \& {Lithwick}, Y. 2013, \apj, 763, 13

\bibitem[{{Zapolsky} \& {Salpeter}(1969)}]{zapolsky+salpeter1969}
{Zapolsky}, H.~S. \& {Salpeter}, E.~E. 1969, \apj, 158, 809

\end{thebibliography}

\clearpage

\clearpage

\end{document}